\documentclass[sn-mathphys,Numbered]{sn-jnl}
\usepackage{upgreek}

\usepackage{graphicx}%
\usepackage{multirow}%
\usepackage{amsmath,amssymb,amsfonts}%
\usepackage{amsthm}%
\usepackage{mathrsfs}%
\usepackage[title]{appendix}%
\usepackage{xcolor}%
\usepackage{textcomp}%
\usepackage{manyfoot}%
\usepackage{booktabs}%
\usepackage{algorithm}%
\usepackage{algorithmicx}%
\usepackage{algpseudocode}%
\usepackage{listings}%

\theoremstyle{thmstyleone}%
\theoremstyle{thmstyletwo}%

\theoremstyle{thmstylethree}%

\begin{document}

\title[Article Title]{Ground Calibration Result of the {\em Lobster Eye Imager for Astronomy}}


\author[1]{\fnm{Huaqing} \sur{Cheng}}

\author*[1,2]{\fnm{Zhixing} \sur{Ling}}\email{lingzhixing@nao.cas.cn}

\author[1,2]{\fnm{Chen} \sur{Zhang}}
\author[3]{\fnm{Xiaojin} \sur{Sun}}
\author[3]{\fnm{Shengli} \sur{Sun}}
\author[1]{\fnm{Yuan} \sur{Liu}}
\author[1]{\fnm{Yanfeng} \sur{Dai}}
\author[1]{\fnm{Zhenqing} \sur{Jia}}
\author[1]{\fnm{Haiwu} \sur{Pan}}
\author[1]{\fnm{Wenxin} \sur{Wang}}
\author[1]{\fnm{Donghua} \sur{Zhao}}
\author[3]{\fnm{Yifan} \sur{Chen}}
\author[3]{\fnm{Zhiwei} \sur{Cheng}}
\author[3]{\fnm{Wei} \sur{Fu}}
\author[3]{\fnm{Yixiao} \sur{Han}}
\author[3]{\fnm{Junfei} \sur{Li}}
\author[3]{\fnm{Zhengda} \sur{Li}}
\author[3]{\fnm{Xiaohao} \sur{Ma}}
\author[3]{\fnm{Yulong} \sur{Xue}}
\author[3]{\fnm{Ailiang} \sur{Yan}}
\author[3]{\fnm{Qiang} \sur{Zhang}}
\author[4]{\fnm{Yusa} \sur{Wang}}
\author[4]{\fnm{Xiongtao} \sur{Yang}}
\author[4]{\fnm{Zijian} \sur{Zhao}}
\author[1,2]{\fnm{Weimin} \sur{Yuan}}


\affil[1]{\orgdiv{Key Laboratory of Space Astronomy and Technology}, \orgname{National Astronomical Observatories, Chinese Academy of Sciences}, \orgaddress{\street{Datun Road 20A}, \city{Beijing}, \postcode{100101}, \country{China}}}

\affil[2]{\orgdiv{School of Astronomy and Space Science}, \orgname{University of Chinese Academy of Sciences}, \orgaddress{\street{Yuquan Road 19A}, \city{Beijing}, \postcode{100049}, \country{China}}}

\affil[3]{\orgdiv{Shanghai Institute of Technical Physics}, \orgname{Chinese Academy of Sciences}, \orgaddress{\street{Yutian Road 500}, \city{Shanghai}, \postcode{1330033}, \country{China}}}

\affil[4]{\orgdiv{Institute of High Energy Physics}, \orgname{Chinese Academy of Sciences}, \orgaddress{\street{Yuquan Road 19B}, \city{Beijing}, \postcode{100049}, \country{China}}}


\abstract{We report on results of the on-ground X-ray calibration of the Lobster Eye Imager for Astronomy (LEIA), an experimental space wide-field (18.6 $\times$ 18.6 square degrees) X-ray telescope built from novel lobster eye mirco-pore optics. 
LEIA was successfully launched on July 27, 2022 onboard the SATech-01 satellite. 
To achieve full characterisation of its performance before launch, a series of tests and calibrations have been carried out at different levels of devices, assemblies and the complete module. 
In this paper, we present the results of the end-to-end calibration campaign of the complete module carried out at the 100-m X-ray Test Facility at the Institute of High-energy Physics, Chinese Academy of Sciences (CAS). 
The Point Spread Function (PSF), effective area and energy response of the detectors were measured in a wide range of incident directions at several characteristic X-ray line energies. Specifically, the distributions of the PSF and effective areas are roughly uniform across the FoV, in large agreement with the prediction of lobster-eye optics. 
The mild variations and deviations from the prediction of idealized, perfect lobster-eye optics can be understood to be caused by the imperfect shapes and alignment of the micro-pores as well as the obscuration of incident photons by the supporting frames, which can be well reproduced by Monte Carlo simulations.
The spatial resolution of LEIA defined by the full width at half maximum (FWHM) of the focal spot ranges from $4$ to $8$ arc minutes with a median of $5.7$ arcmin. 
The measured effective areas are in range of $2-3~\rm{cm^2}$ at $\sim$1.25 keV across the entire FoV, and its dependence on photon energy is also in large agreement with simulations. 
The gains of the CMOS sensors are in range of $6.5-6.9~{\rm eV/DN}$, and the energy resolutions in the range of $\sim120 - 140$ eV at $1.25$ keV and $\sim170-190$ eV at $4.5$ keV. 
These calibration results have been ingested into the first version of calibration database (CALDB) and applied to the analysis of the scientific data acquired by LEIA.
This work paves the way for the calibration of the Wide-field X-Ray Telescope (WXT) flight model modules of the Einstein Probe (EP) mission.
}


\maketitle

\section{Introduction}\label{sec:intro}

As a novel X-ray focusing imaging technique, lobster eye micro-pore optics (MPO) provide a promising tool for wide-field X-ray monitoring \cite{1979ApJ...233..364A, 1992SPIE.1546...41F, 1998ExA.....8..281W, 2002SPIE.4497..115F, 2016SPIE.9905E..1YW}. 
The Lobster Eye Imager for Astronomy (hereafter LEIA)
is the very first of its kind with a considerably large field of view ever flown (see \cite{2023LingZXRAA} for a detailed account of LEIA).
As the pathfinder of the Einstein Probe (EP) mission\footnote{The Einstein Probe is a time-domain X-ray mission of the Chinese Academy of Sciences (CAS), in collaboration with the European Space Agency (ESA), the Max-Planck Institute for Extraterrestrial Physics (MPE) and the France Space Agency (CNES).} \cite{2015arXiv150607735Y, 2018SPIE10699E..25Y, 2022hxga.book...86Y}, 
the main goal of LEIA is to demonstrate the in-orbit performance of the MPO and CMOS devices, from which the wide-field X-ray telescope (WXT) of EP is built. 
LEIA was launched successfully on July 27, 2022, onboard the CAS's SATech-01 satellite.
SATech-01 has a sun-synchronous orbit of 500 km in height, with a period of 95 minutes. 
LEIA has been operating in orbit for more than one year so far. 
Its first-light results have been published in \cite{2022ZhangChenApJL}.

\begin{figure*}
    \centering
    \includegraphics[width=0.6\textwidth]{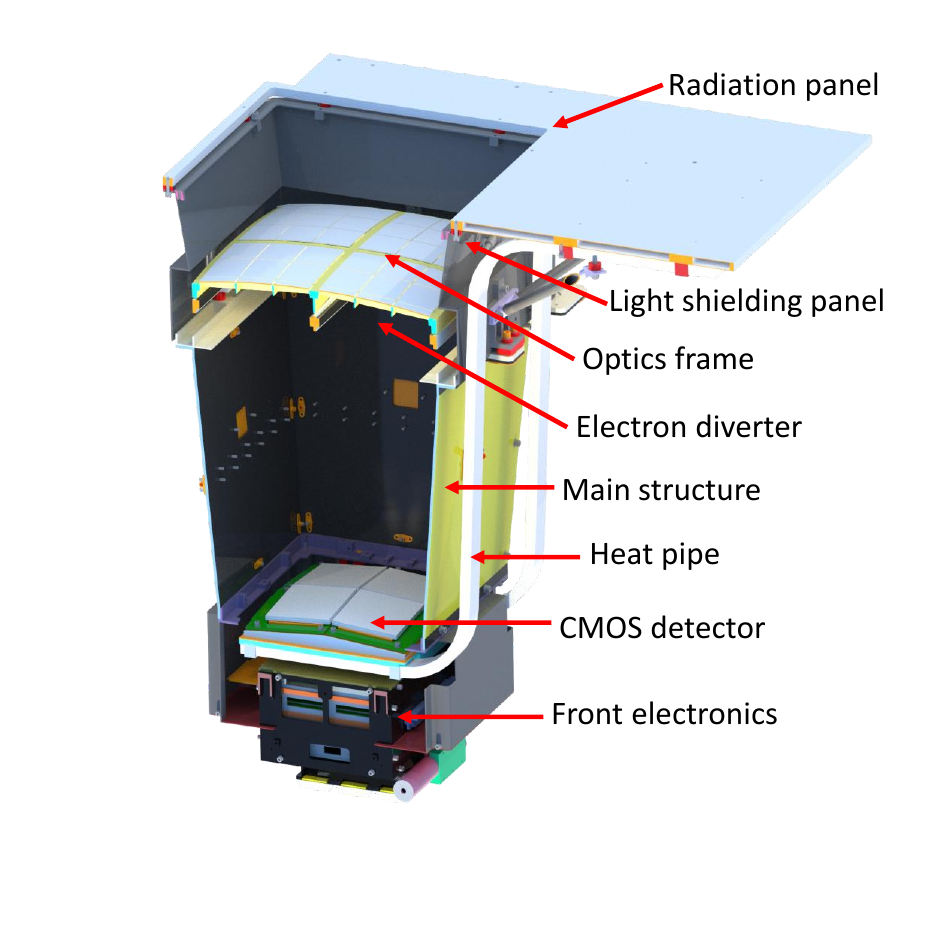}
    \caption{Schematic figure of the {\em LEIA} instrument (from Fig.1 of \cite{2023LingZXRAA}).}
    \label{fig:leia_schematic}
\end{figure*}

The LEIA instrument is a qualification model of one of the twelve identical modules of EP WXT, developed jointly by the National Astronomical Observatories of CAS (NAOC) and the Shanghai Institute of Technical Physics (SITP). 
The design of its optics and layout can be found in Fig. 1 of \cite{2022ZhangChenApJL}.
The schematic plot of the instrument is shown in Fig. \ref{fig:leia_schematic} and the specifications are listed in Table \ref{tab:leia_specification}.
LEIA has a field of view (FoV) as large as 18.6 $\times$ 18.6 square degrees.
For an ideal lobster eye telescope, there is almost no vignetting effect across the FoV except for the edges in theory; in practice, however, there is non-uniformity within the FoV caused by a few factors, such as defects of the MPO devices and obscuration of light by the MPO supporting structure of the mirror frame (see \cite{2022ZhangChenApJL}). 
The telescope is composed of mainly two key components: a lobster eye focusing mirror assembly (MA) built from 36 MPO plates, and a focal-sphere detector module made of four large-format scientific back-illuminated CMOS sensors (labeled as CMOS 1-4) \cite{2022PASP..134c5006W,wu2023}. 
Each of the CMOS sensors, subtending one quadrant of the entire FoV, has 4096 $\times$ 4096 pixels, with a pixel size of 15 $\mu$m by 15 $\mu$m. 
The top of the sensor is coated with 200 nm aluminium to block optical and ultraviolet light. 
We also note that this is, to the best of our knowledge, the first application of this kind of CMOS sensor as X-ray detector in orbit.

\begin{table}[h!]
\centering
\caption{Specifications of the LEIA instrument.}
\begin{tabular}{ l | l }
\hline
Orbit & 500 km  \\ \hline
Launch date & July 27 2022 \\ \hline
Optic & 36 MPO plates \\ \hline
Detector & Four 6 cm $\times$ 6 cm CMOS sensors \\ \hline
Field of View & 346 square degrees  \\ \hline
Focal length & 375 mm \\ \hline
Angular resolution & 4 $\sim$ 8 \ arcmin (FWHM) \\ \hline
Energy band & 0.5 -- 4 keV  \\ \hline
Energy resolution &  125 -- 137 eV at 1.25 keV  \\ \hline
Time resolution & 50 ms  \\ \hline
Weight and Power  &  53 kg, 85 W  \\ \hline
\end{tabular}
\label{tab:leia_specification}
\end{table}

Before the launch of LEIA, extensive experiments were carried out to characterize the instrumental performance and calibrate the properties.
These tests were done on different levels: device, 
assembly and the complete instrument. 
Specifically, for the mirror assembly, the following tests were adopted.  
First, before mounting onto the MA, each of the MPO plates was individually measured in order to derive its parameters and to assess its optical performance.
Secondly, the MA was first tested to measure its imaging point spread function (PSF) at the X-ray Imaging Beamline (XIB) at NAOC \cite{2012SPIE.8443E..3XZ}. It was later calibrated mainly for its effective area (as well as PSF independently) at the Panter X-ray Test Facility of MPE, Germany\cite{2005ExA....20..405F, 2019SPIE11119E..16B}. The calibration results at MPE/Panter have been published \cite{2023SPIE12777E..3FR}.
Meanwhile, the performances of the CMOS detectors were tested separately at NAOC, whose results will be presented elsewhere (Ling et al. in preparation).
Then the MA and the detector module were integrated together to build the complete module of the LEIA instrument. Finally, the complete instrument was calibrated at the 100-m X-ray Test Facility of the Institute of High Energy Physics (IHEP) \cite{2023ExA....55..427W} in November 2021. 
In this paper, we presents the results of the final end-to-end calibration campaign carried out at IHEP, with the main focus on the PSF, effective area and energy response. The calibration results have been ingested into the calibration database (1st version) and applied to the analysis of LEIA data.
The paper is organised as follows. The experiment setup, procedure and data processing are described in Section \ref{sec:overview}. The calibration results are presented in Section \ref{sec:calib_results} and the discussions are presented in Section \ref{sec:discuss}. We summarize the calibration result in Section \ref{sec:summary}.

\section{Overview of the calibration experiment}
\label{sec:overview}

\subsection{Experiment setup}
\label{sec:experiment_setup}

The 100-m X-ray Test Facility (100XF) \cite{2023ExA....55..427W} is a dedicated facility built by IHEP, CAS, to calibrate X-ray telescopes. 
The main components of the 100XF are the large instrument chamber (8-m long and 3.4-m in diameter), X-ray sources, a 100-m vacuum tube, pump stations, movable stages, and standard X-ray detectors. An X-ray beam is generated by an X-ray source at a distance of 100 meters from the chamber and has a beam size of 0.6 meters in diameter. The beam is approximately quasi-parallel with a divergence angle of $<9$ arcmin at the exit end of the chamber\cite{2023ExA....55..427W}. We note that the LEIA instrument was $\sim1.4$ millimeter out of focus due to the finite distance of the point source.

\begin{figure*}
    \centering
    \includegraphics[width=0.6\textwidth]{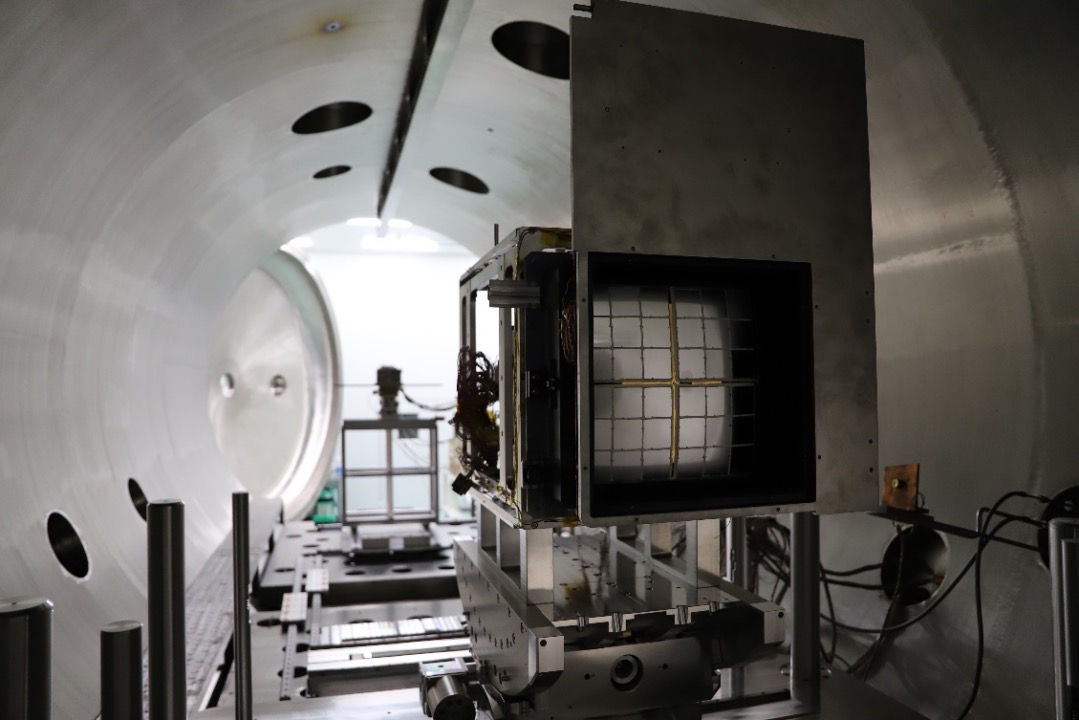}
    \caption{The LEIA instrument in the large chamber of the 100XF at IHEP.}
    \label{fig:qm4_at_ihep}
\end{figure*}

The LEIA instrument was first degassed in a small vacuum tank (with a vacuum degree of $10^{-3}~{\rm Pa}$) for 3 days. Then it was mounted in the large instrument chamber (Fig. \ref{fig:qm4_at_ihep}) and vacuumed to a higher degree of $\sim10^{-5}~{\rm Pa}$. During the experiments, the working temperature of the mirror assembly was kept at $\sim 20$ degrees and that of the CMOS detectors at  $-30$ degrees. A multi-target electron impact X-ray source was utilized to generate X-ray beams at a number of photon energies. The X-ray source targets used were magnesium (Mg), titanium (Ti), silicon dioxide (SiO$_2$), silver (Ag) and copper (Cu). A standard silicon drift detector (SDD), whose quantum efficiency (QE) has been calibrated by a standard PNCCD, was installed beside the mirror assembly to monitor the absolute count rate of the X-ray beam. 

\begin{figure*}
    \centering
    \includegraphics[width=0.6\textwidth]{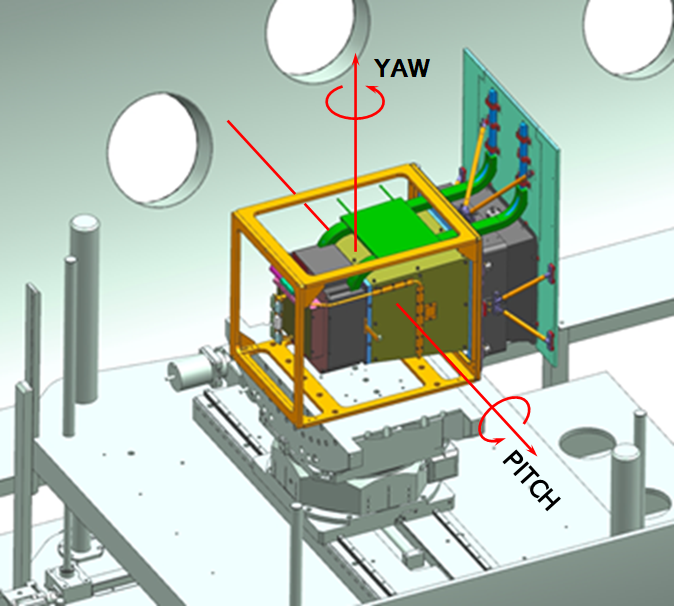}
    \caption{Schematic diagram for the movable stage. The attitude of the instrument can be adjusted by rotating the two axes (corresponding to different combinations of pitch and yaw angles).}
    \label{fig:qm4_rotate}
\end{figure*}

\begin{figure*}
    \centering
    \includegraphics[width=0.6\textwidth]{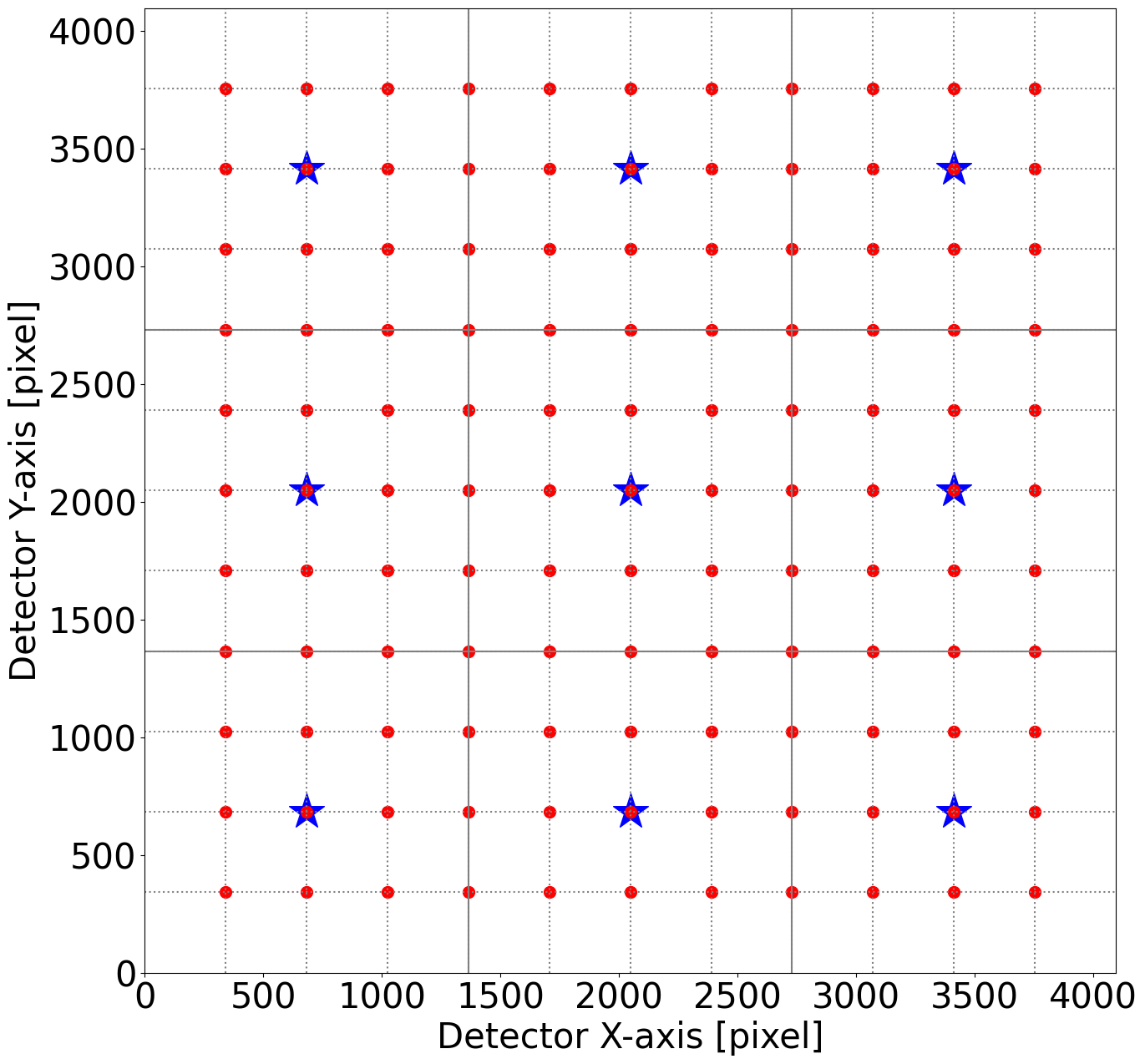}
    \caption{The diagram of the scan array for the mapping of the PSF and effective area across a quadrant of the FOV. The red dots contribute to the `fine scan' array of $11\times11$ test points on the detector corresponding to 121 incident directions sampled within the FoV quadrant, adopted in the experiment using the Mg source. The blue pentagrams contribute to the `chip center scan' array of $3\times3$ test points on the detector corresponding to 9 incident direction sampled, adopted in the experiment of SiO$_2$, Ti and Ag sources. See text for more details.}
    \label{fig:scanarray}
\end{figure*}

To investigate the imaging quality (PSF) and the effective area across the FoV, the module was mounted on a movable stage in the chamber (Fig. \ref{fig:qm4_rotate}). The attitude of the instrument can be adjusted by changing the pitch and yaw angles. 
Fig. \ref{fig:scanarray} shows, as an example, the sampling array of the test points on one of the four CMOS detectors, which corresponds to a quadrant of the entire FOV. 
The array consists of 11 $\times$ 11 points on the detector plane (red dots), the theoretical positions of the central focal spots of the corresponding 121 light incident directions within the FoV quadrant. 
They define the `fine scan' array of test points.
In addition, we also define a sub-grid of 3$\times$3 points across the CMOS (blue pentagram in Fig. \ref{fig:scanarray}), which are the theoretical positions of the focal spots of the light incident directions passing through the center of the 3$\times$3 MPO plates.   
These sampling positions (directions) are referred to as the `chip centre scan' array. 
For the measurements using the Mg source at 1.25 keV, the `fine scan' array directions were sampled. For the measurements using the Ti, SiO2 and Ag sources (at 0.53, 1.74, 2.98 and 4.51 keV, respectively), the `chip centre scan' array directions were sampled. Finally, for the measurement using the Cu source at 0.93 keV, only the direction passing through the CMOS chip center was sampled.

\subsection{Experiment procedure}
\label{sec:experiment_schedule}

Before illuminating the instrument with the X-ray beam, several dark images (i.e. the bias map) were acquired and analysed to derive bias residual and to identify bad pixels. During the calibration procedure the dark images were taken for several times to diagnose the stability of the detectors. No prominent changes were found, however.

The end-to-end calibration campaign was performed from November 14 to 18, 2021.
The LEIA telescope was illuminated by X-ray beams with several photon energies and at different incident directions that correspond to the focal spot positions defined by the grid spots of a number of combination as explained above and in Fig. \ref{fig:scanarray}. 
The logs of the calibration experiment are summarized in Table \ref{tab:leia_scan_schedule}. 
Note that the measurements in the largest number of the test directions (121 points for each CMOS and hence 484 for all the four CMOS detectors) were made at 1.25 keV only, due to the limited time available for the calibration campaign. The exposure times for different directions at different photon energies were optimized in consideration of the strength of the X-ray light and the overall time length of the calibration campaign. For all the measurements the statistical uncertainties were estimated to be negligible.

\begin{table}[htbp]
\caption{Log of the calibration experiment.}
\begin{tabular}{ccccc}
\toprule%
\centering
Date & Target & Line energies & Sampling points per CMOS & Exposure (s) \\ \hline
2021.11.14 & SiO2 & 525 eV, 1740 eV & $3\times3$ & 600 \\
2021.11.15 & Ag & 2980 eV & $3\times3$ & 500 \\
2021.11.16 & Ti & 4511 eV, 4932 eV & $3\times3$ & 500 \\
2021.11.17-11.18 & Mg & 1254 eV &$11\times11$ & 80 \\
2021.11.18 & Cu & 930 eV &$1$ & 2000 \\ 
\botrule
\end{tabular}
{{\rm Notes:} A summary of the experimental setup for different X-ray sources, including the date, target, test energies of characteristic X-ray lines, sampling points on each CMOS detector and the exposure time of each test point.}
\label{tab:leia_scan_schedule}
\end{table}

\subsection{Data processing}
\label{sec:data_process}

X-ray photon events were extracted and readout from the LEIA detectors (see \cite{2022PASP..134c5006W} for a detailed account of the extraction of X-ray photon events by CMOS detectors). Only signals with the amplitude larger than the low-energy threshold (about 350 eV in the normal working mode and adjustable) are saved during the image reading out.

Each of the extracted X-ray events was assigned with a summed pulse-height amplitude (PHA) and a grade of the event pattern. X-ray events with the single, double, triple, and quadruple grades were selected, among all the events, to generate high-level products such as images, spectra, and light curves. These products form the basis for the investigation of the PSF, effective area, and detector energy responses, as described in below. 

\section{Calibration Result}
\label{sec:calib_results}

For an imaging telescope, the essential properties to be calibrated includes the PSF, effective area as well as their variations across the whole field of view (vignetting), the energy response function of the detectors, and other items related to the scientific products. In this section, we present the calibration results of these properties. 

\subsection{Point Spread Function}
\label{sec:psfscan}

Compared to Wolter-I telescopes, lobster eye optics have a complicated point spread function. It comprises a bright central focal spot and cruciform arms. The focal spot is contributed by photons that undergo two or even times of reflections off two adjacent (perpendicular) walls of micro-square pore channels, whereas the cruciform arms are formed by those that undergo an odd number of reflections. The PSF of LEIA was measured at a grid of different incident angles within the FoV and at several energies of X-ray characteristic lines.

\subsubsection{PSF across the FoV}
\label{sec:psf_dif_direction}

\begin{figure*}
\centering
\includegraphics[width=0.45\textwidth]{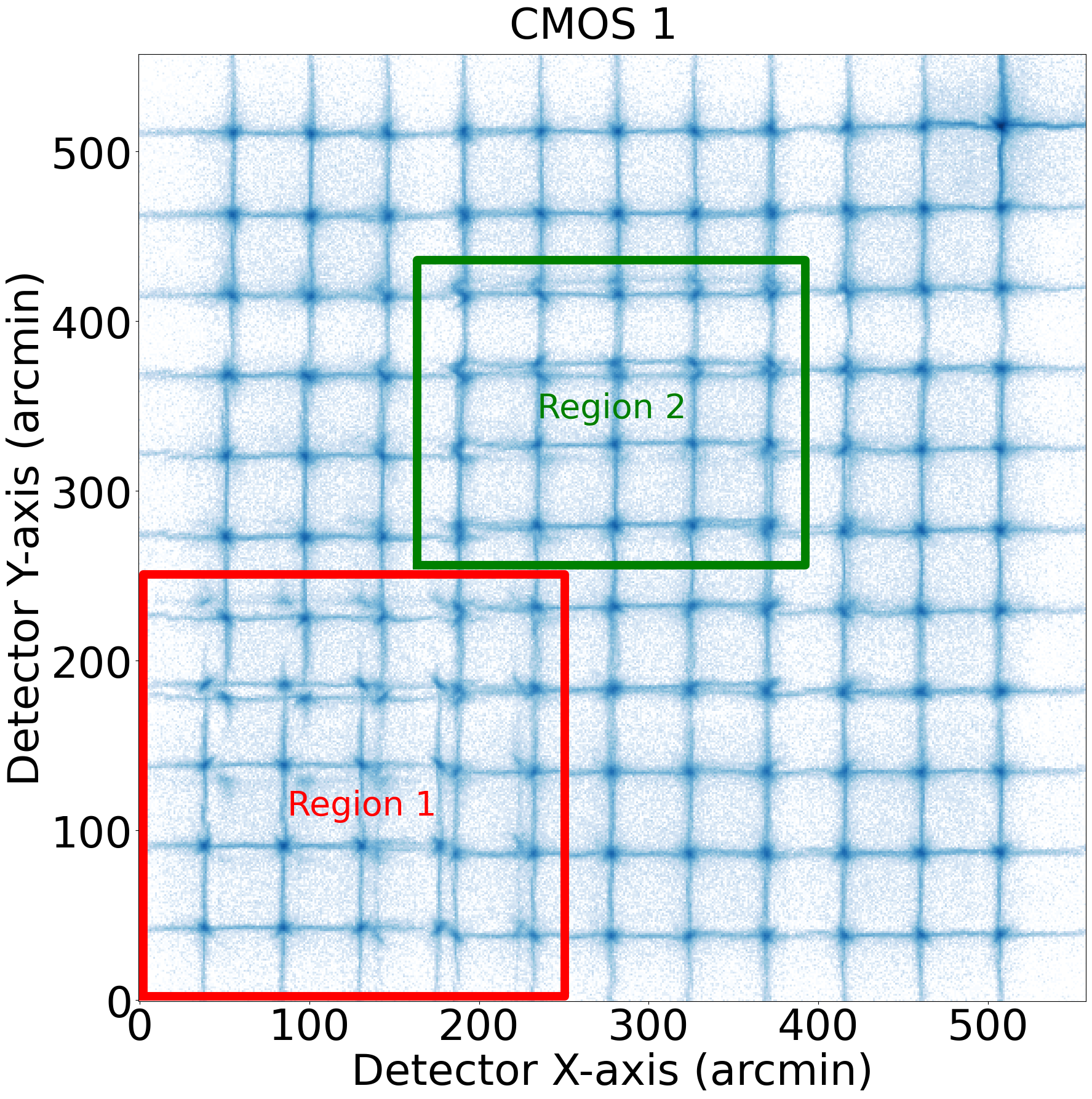}
\includegraphics[width=0.45\textwidth]{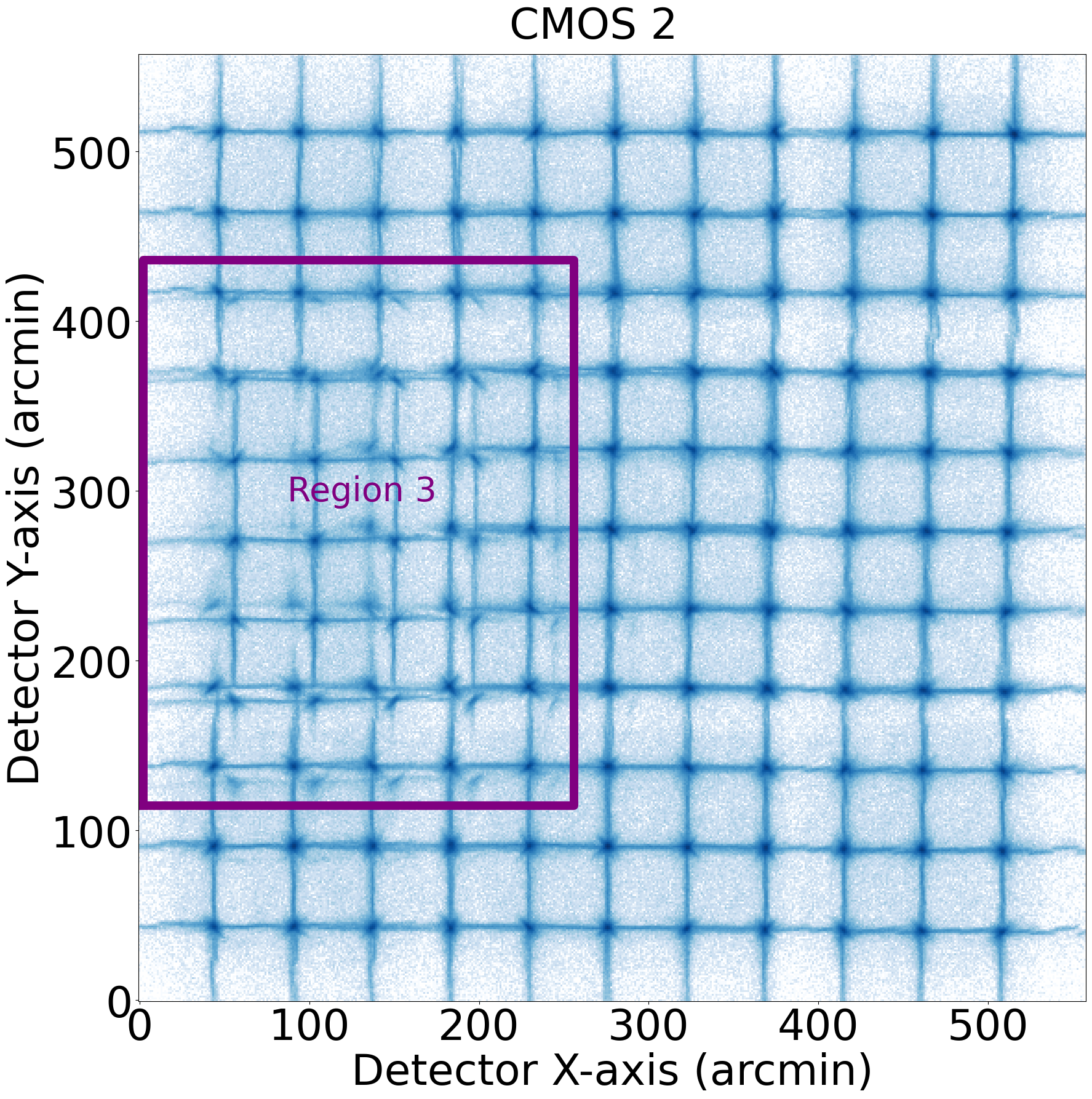}
\includegraphics[width=0.45\textwidth]{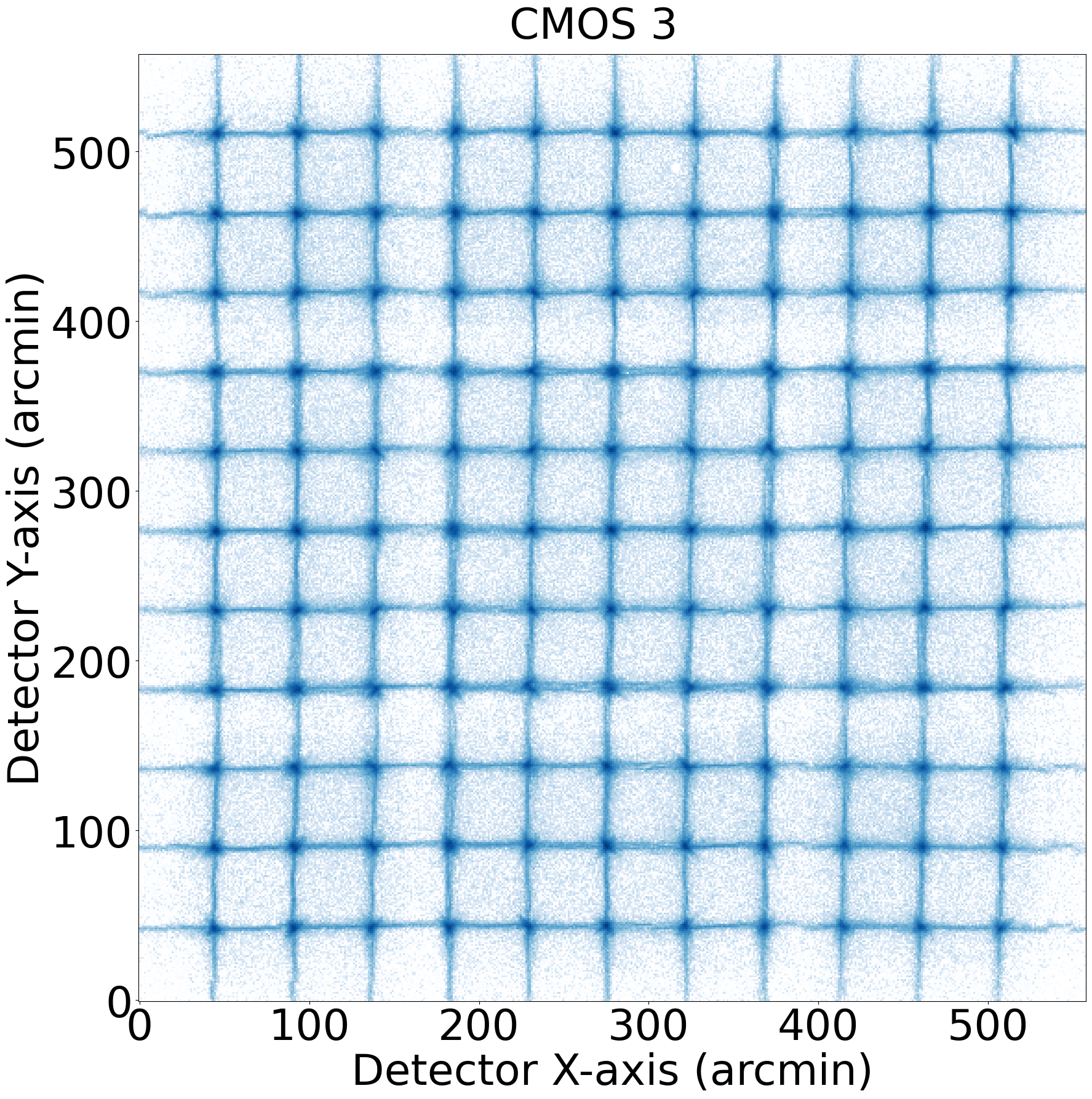}
\includegraphics[width=0.45\textwidth]{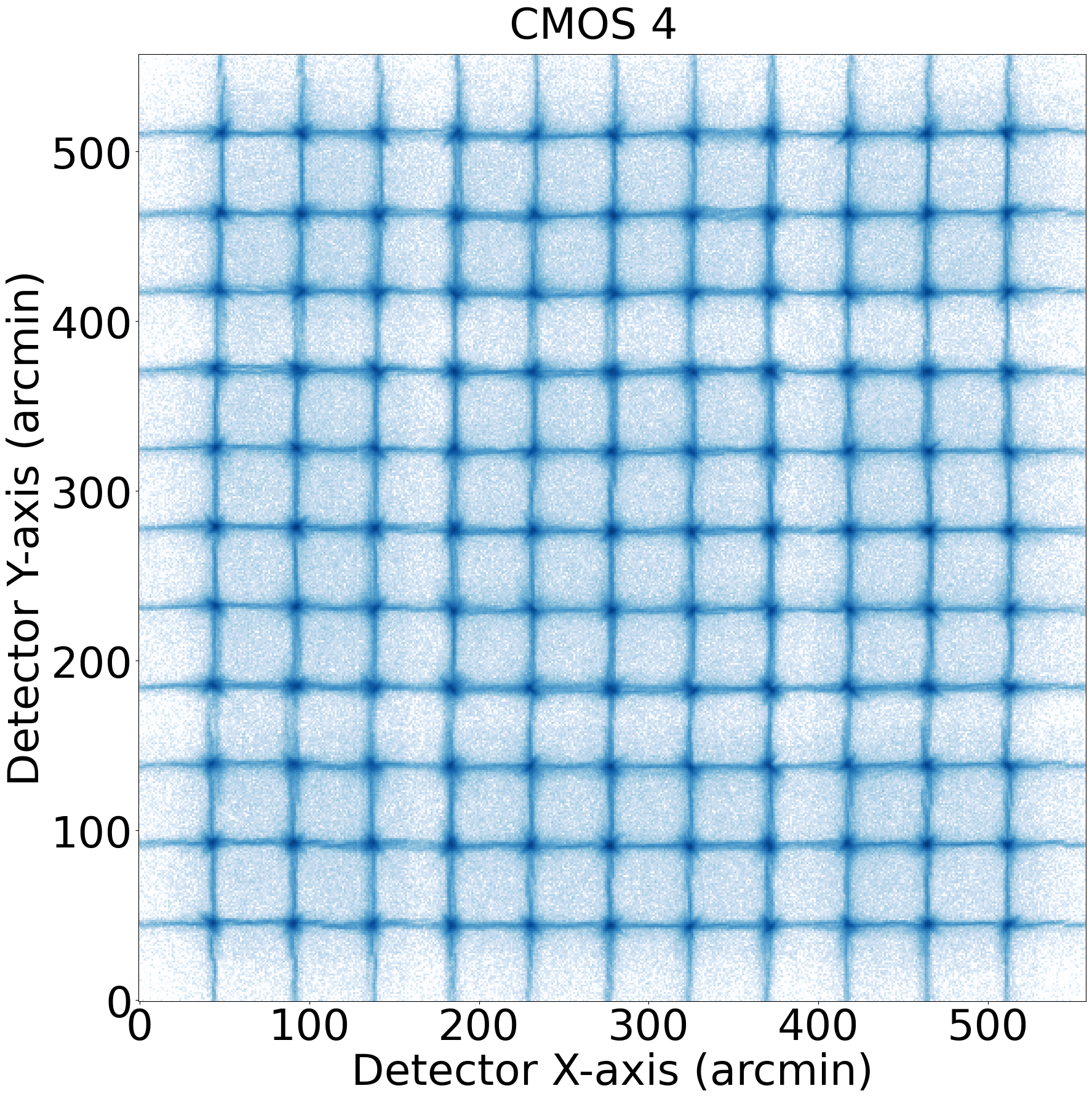}
\caption{\label{fig:psfscan_mgk}A mosaic of images of the PSF obtained at 1.25 keV across the entire FoV of LEIA taken with the complete module at 100XF. For CMOS 1 and CMOS2, there are three confined regions showing PSF misalignment (Region 1, Region 2 on CMOS 1, Region 3 on CMOS2). See text for details.}
\end{figure*}

\begin{figure*}
    \centering
    \includegraphics[width=\textwidth]{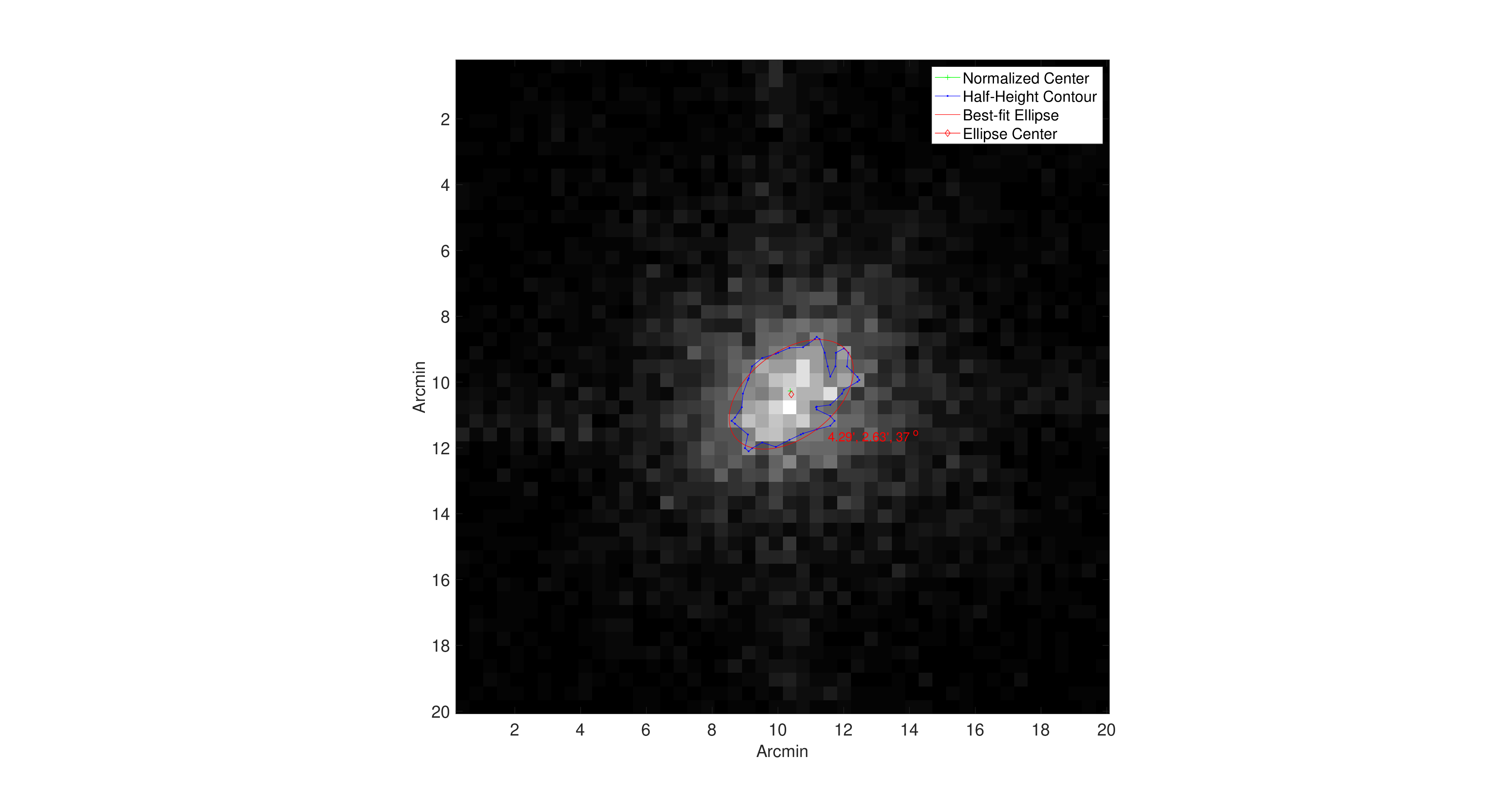}
    \caption{An example of the elliptical fitting for the focal spot. The green cross is the barycenter of the focal spot region, the blue solid line is the half-height contour line, and the red solid line is the best-fit elliptical function of the half-height contour with the diamond denoting the ellipse center. In this case, the lengths of the long axis and short axis of the best-fit elliptical function are $\sim4.29$ arcmin and $\sim2.63$ arcmin respectively, and the position-angle is $\sim37$ degrees.}
    \label{fig:psf_fitting_ellipse}
\end{figure*}

\begin{figure*}
    \centering
    \includegraphics[width=\textwidth]{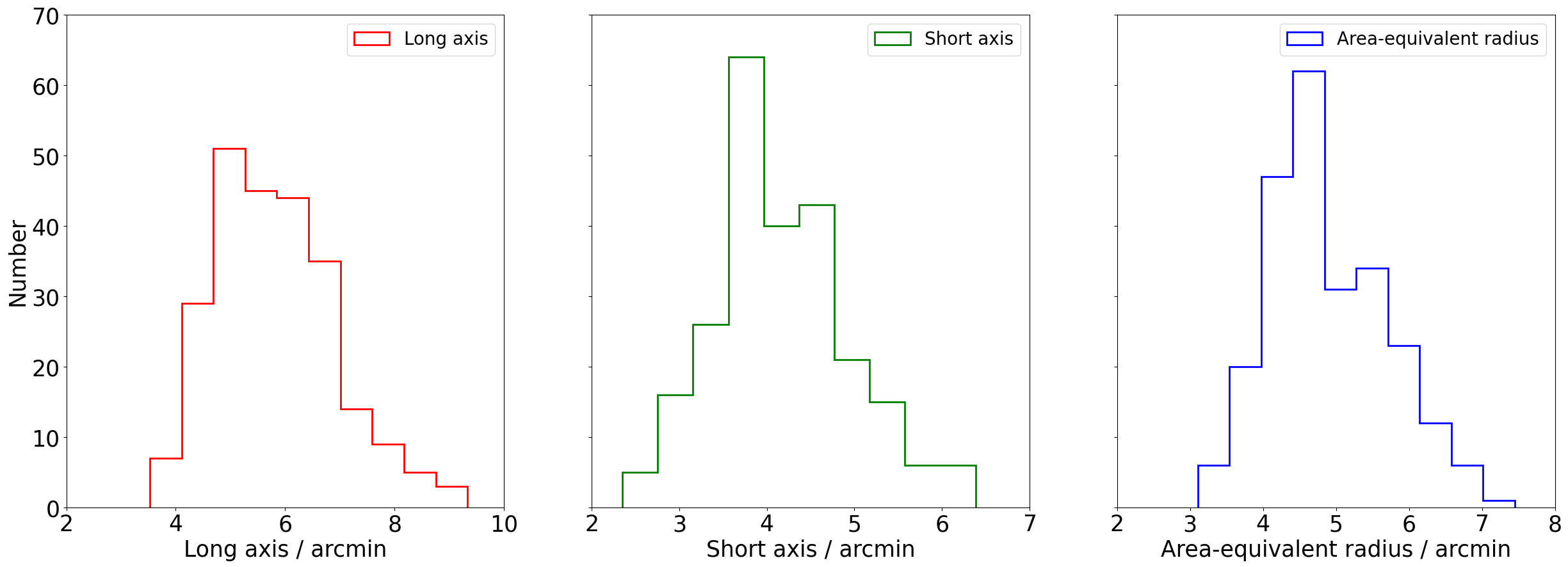}
    \caption{The distribution of all three measures of the FWHM (long axis, short axis and area-equivalent radius) of the sampled PSFs within the FoV quadrants of CMOS 3 and CMOS 4.}
    \label{fig:psf_fitting_cdf}
\end{figure*}

For each quadrant of the FoV (subtended by one CMOS sensor), exposures of 80 seconds each were taken toward 11$\times$11 sampling directions at the energy of the Mg K$\alpha$ line (1.25 keV). The mosaics of the X-ray images of the PSF across the whole FoV is shown in Fig. \ref{fig:psfscan_mgk} for all the four CMOS sensors. The characteristic cruciform shapes of the PSFs are found to be largely consistent among most of the sampled directions. 
This basically confirms the homogeneity of the imaging property across essentially the FoV predicted for the lobster-eye optics\cite{1979ApJ...233..364A, 2014SPIE.9144E..4EZ, 2016SPIE.9905E..1YW}.
For two quadrants of the FoV (corresponding to detectors CMOS 3 and CMOS 4), the central focal spots and arms are very well aligned, indicating that the anterior MPO plates of the mirror assembly are precisely mounted (for an ideal lobster eye mirror the MPO should form part of a perfect sphere). However, for the other two quadrants of the FOV (corresponding to CMOS1 and CMOS 2), a considerable fractions of the PSFs are misaligned in three separate regions, two on CMOS 1 and one on CMOS 2. It can be seen that most of the PSF in the misalignment regions are actually split into two or even three parts. This issue will be discussed in Section \ref{sec:psf_misalignment}.

The structures of the PSFs were investigated in detail.
It is found that the actual shape of the central focal spot is elongated from a circle to an ellipse (as predicted by simulations, \cite{2021OptCo.48326656L, 2022PASP..134k5002L}) as shown in Fig. \ref{fig:psf_fitting_ellipse}. The barycenter of the PSF (the `normalized center' in the figure), as well as the contours, can be derived for the focal spot region. An elliptical function is then used to fit the half-height contours. The long and short axes of the ellipse, as well as an `area-equivalent' radius (defined as the square root of the product of the long and short axes, i.e. a measure of the area), were measured, which are considered to be representative of the full width at half maximum (FWHM) of the focal spot. In this paper, the length of the long axis is defined as the spatial resolution of the PSF as it represents the limit of the resolving capability. Fig. \ref{fig:psf_fitting_cdf} shows the distribution of all three measures of the FWHM for the 242 test directions from CMOS 3 and CMOS 4 that do not suffer from defects of the PSF misalignment. 
It can be seen that the spatial resolutions of the PSF of LEIA thus defined range from 4 arcmin to 8 arcmin.

\subsubsection{Dependence on photon energy}
\label{sec:psf_on_energy}

\begin{figure*}
    \centering
    \includegraphics[width=0.3\textwidth]{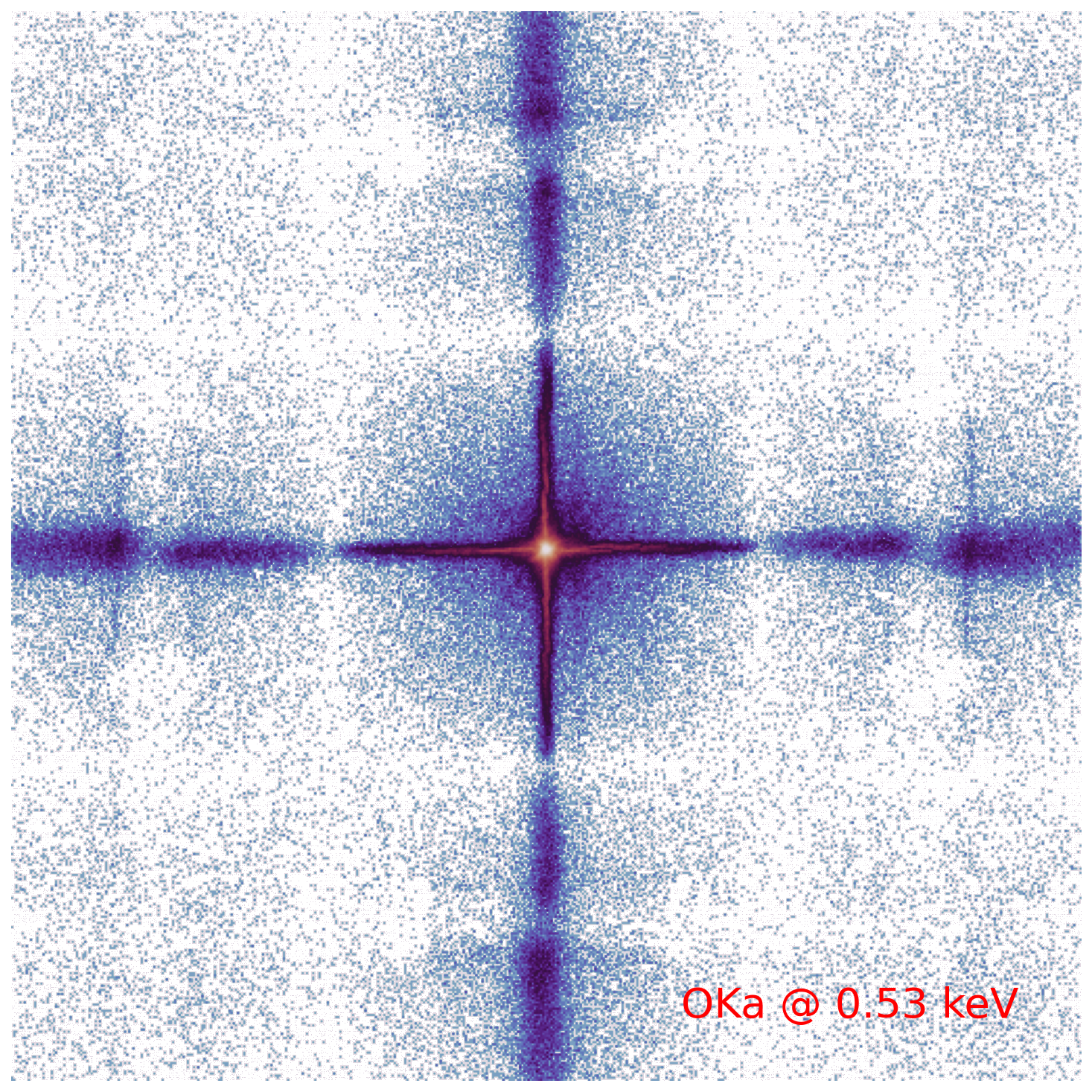}
    \includegraphics[width=0.3\textwidth]{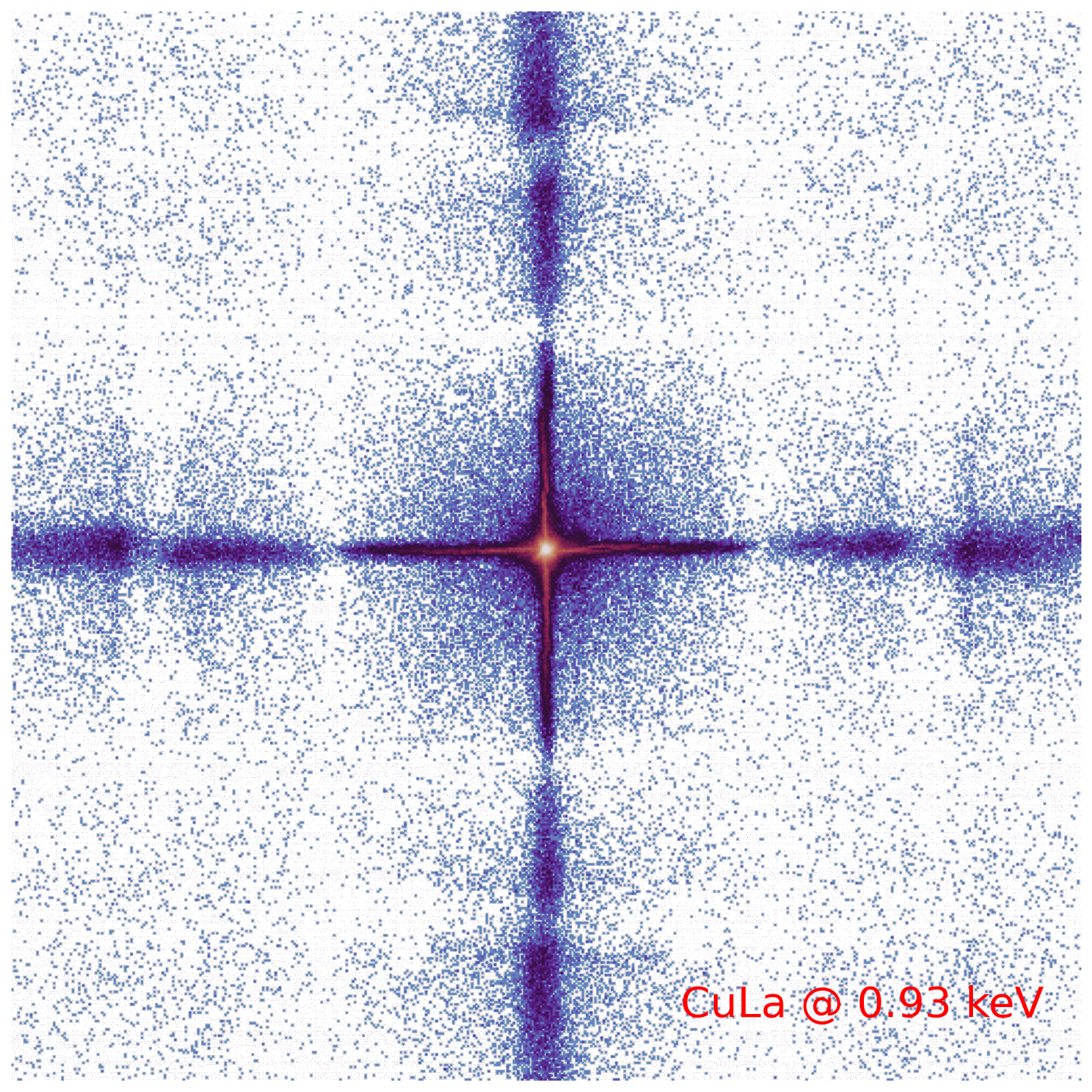}
    \includegraphics[width=0.3\textwidth]{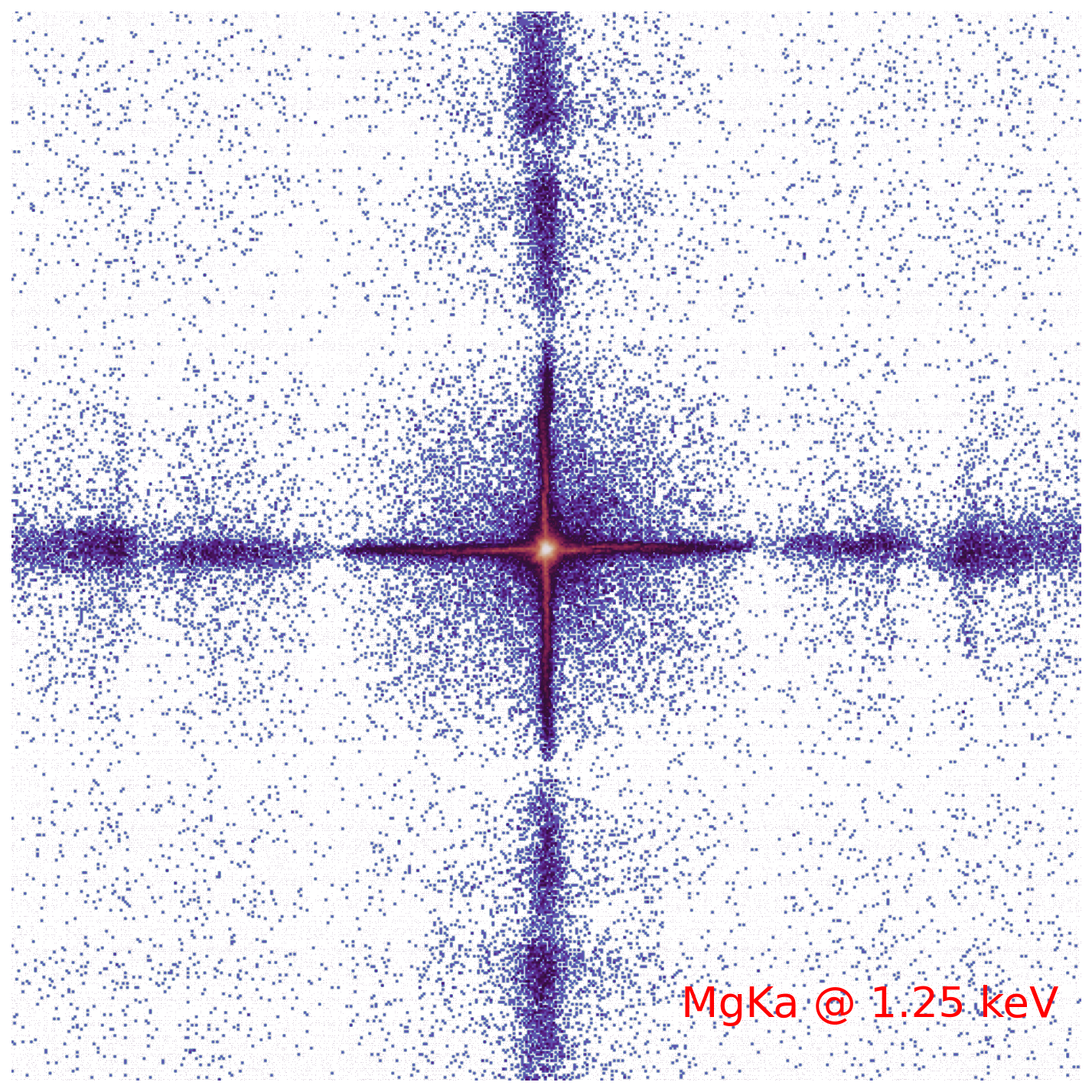}
    \includegraphics[width=0.3\textwidth]{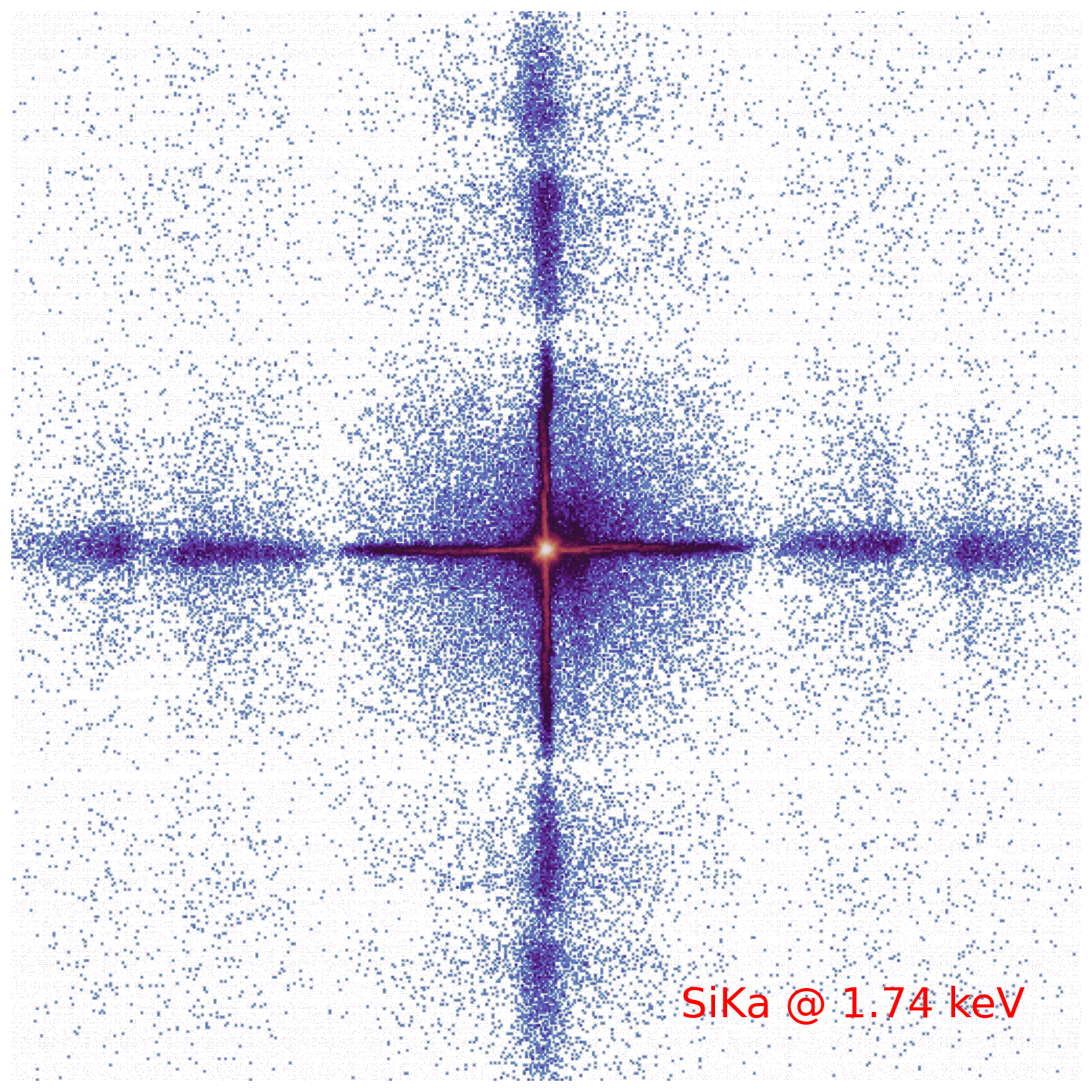}
    \includegraphics[width=0.3\textwidth]{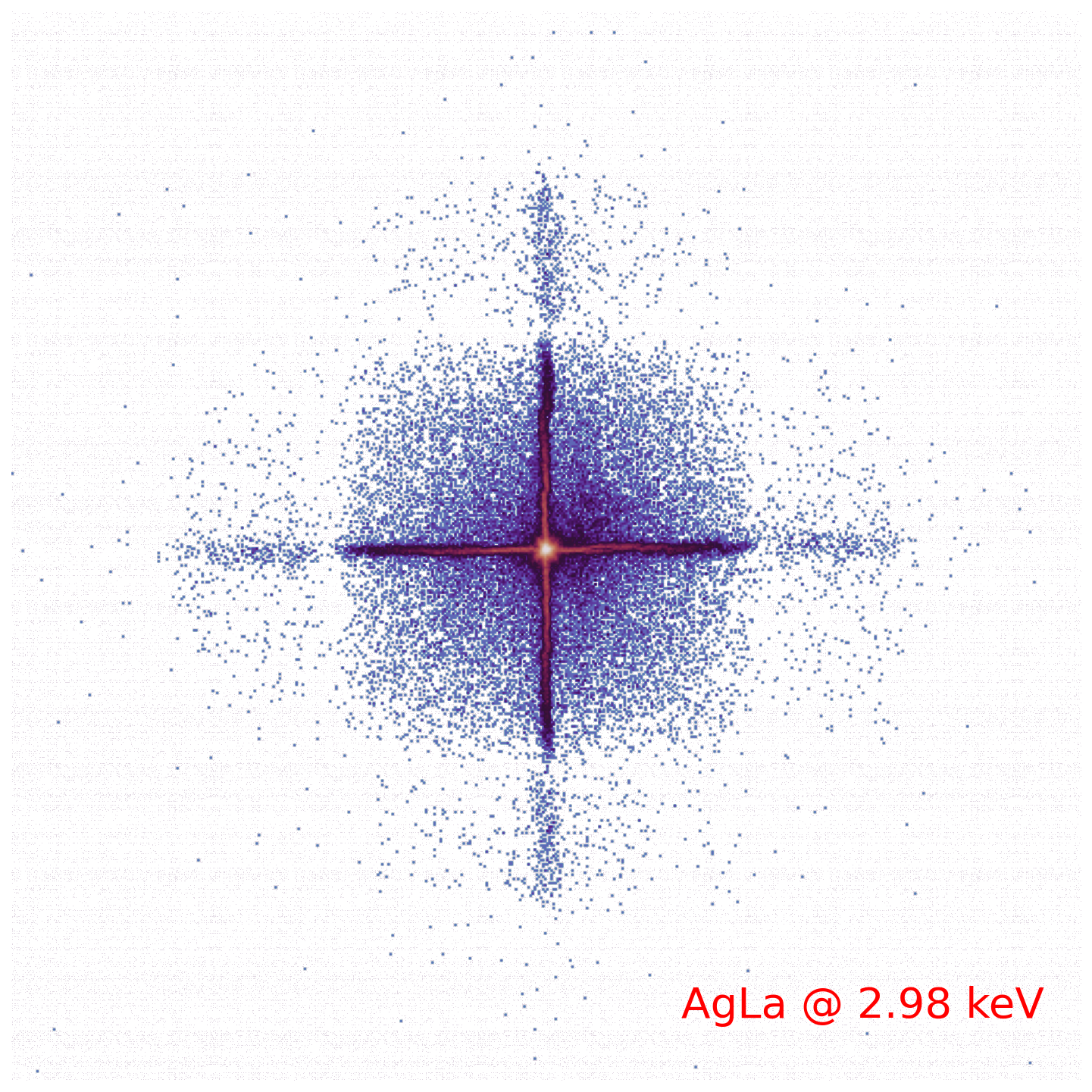}
    \includegraphics[width=0.3\textwidth]{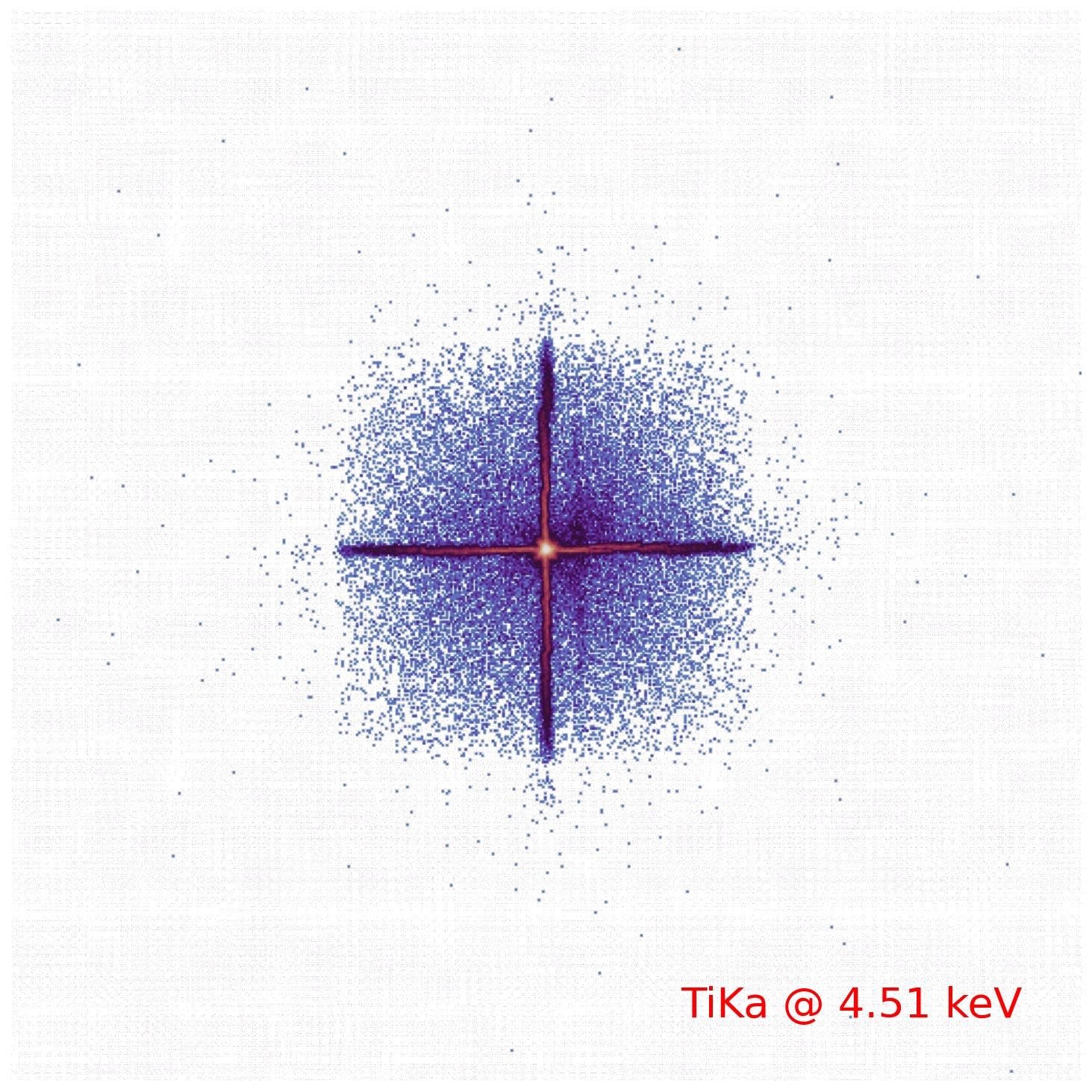}
    \caption{Images of the measured PSFs in the direction along the center of CMOS 4, taken at the line energies of O K$\alpha$, Cu L$\alpha$, Mg K$\alpha$, Si K$\alpha$, Ag L$\alpha$ and Ti K$\alpha$.}
    \label{fig:psf_dif_en}
\end{figure*}

The PSFs were also measured for the images taken at several other energies of the incident X-rays. Limited by the overall time duration of the calibration campaign, only an array of 9 ($3\times 3$)  directions within the FoV of each CMOS (i.e. a $3\times 3$ array of positions on the CMOS detectors) were sampled at the line energies of O K$\alpha$ (525 eV), Si K$\alpha$ (1740 eV), Ag L$\alpha$ (2980 eV) and Ti K$\alpha$ (4510 eV), and only the direction along the center of the CMOS was sampled at the Cu L$\alpha$ line energy (930 eV). 
As an example, Fig. \ref{fig:psf_dif_en} shows the 
measured PSFs in the directions along the center of CMOS 4 at the above five energies plus the 1.25 keV of Mg K$\alpha$ line.
It is found that lower energy photons spread out over a larger area than higher energies photons. The cruciform arms contributed by photons that undergo three times of reflections can hardly be recognised at the energies of Ag L$\alpha$ and Ti K$\alpha$ lines. These properties are consistent with the results of the independent calibration of the mirror assembly carried out at MPE/Panter \cite{2023SPIE12777E..3FR}. 

\subsection{Effective Area}
\label{sec:effarea}

\begin{figure*}[htbp]
    \centering
    \includegraphics[width=0.75\textwidth]{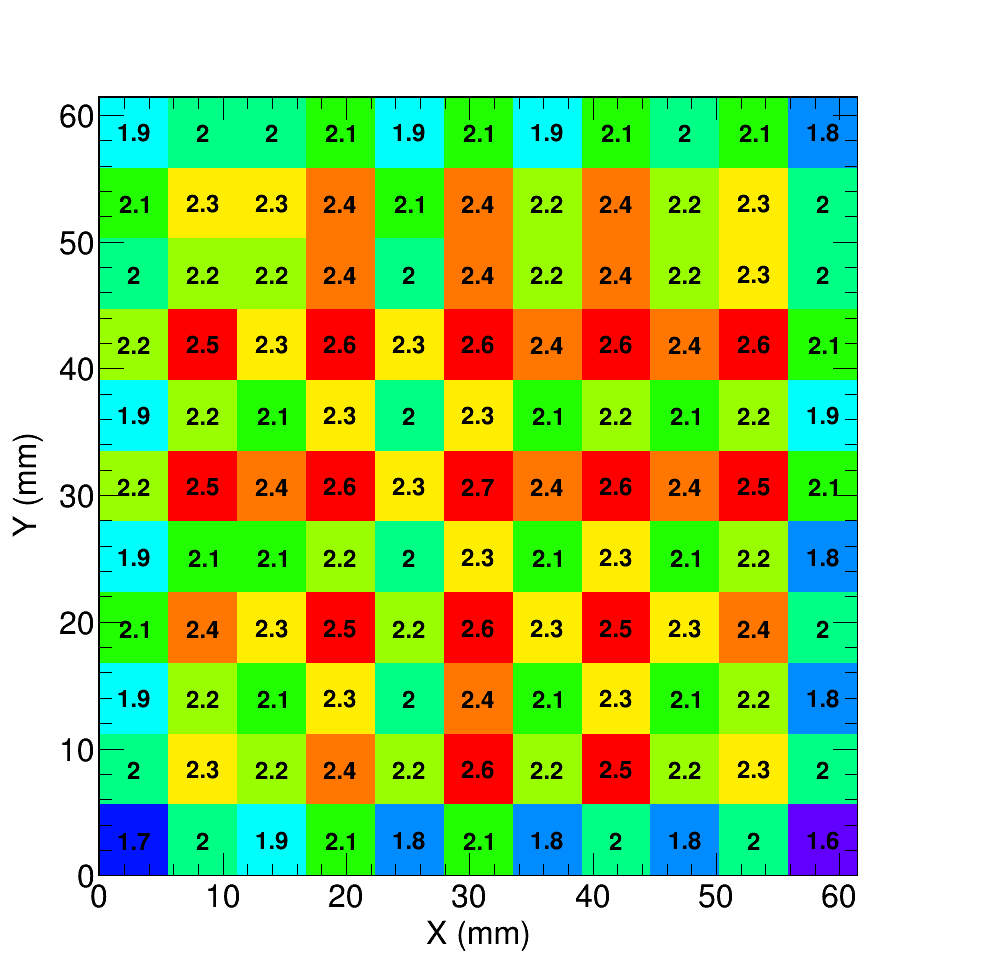}
    \caption{Distribution of the effective area (for the focal spot only) at the energy of Mg K$\alpha$ line (1.25 keV) within the FoV quadrant of CMOS 4.}
    \label{fig:vignetting_figure}
\end{figure*}

\begin{figure*}[htbp]
    \centering
    \includegraphics[width=0.45\textwidth]{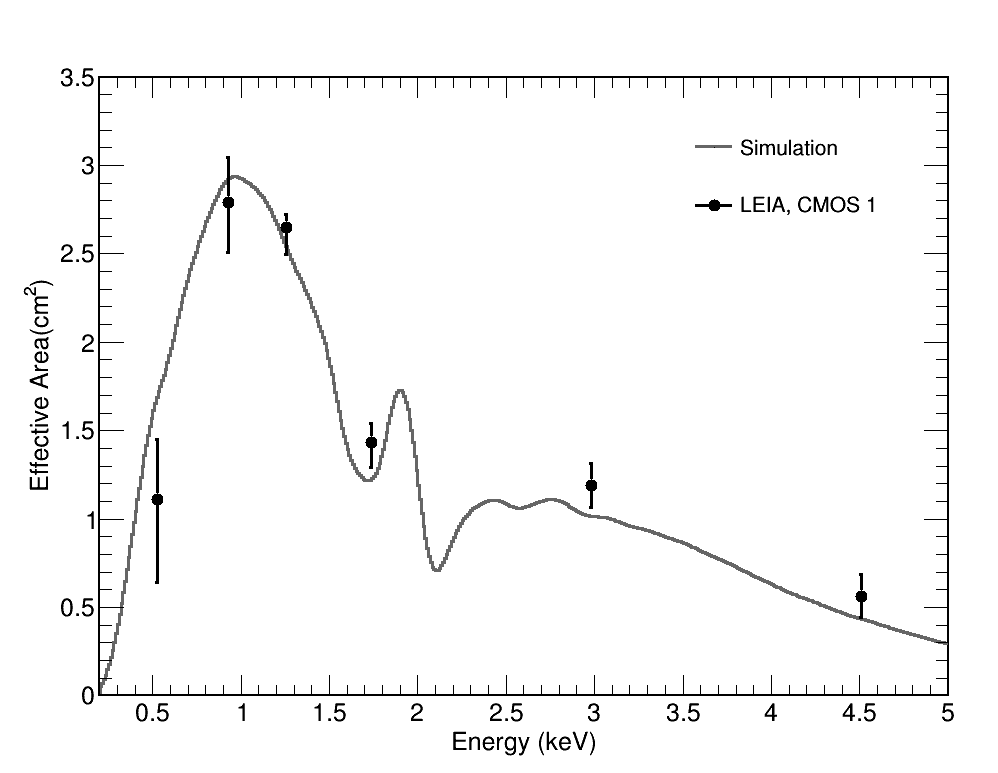}
    \includegraphics[width=0.45\textwidth]{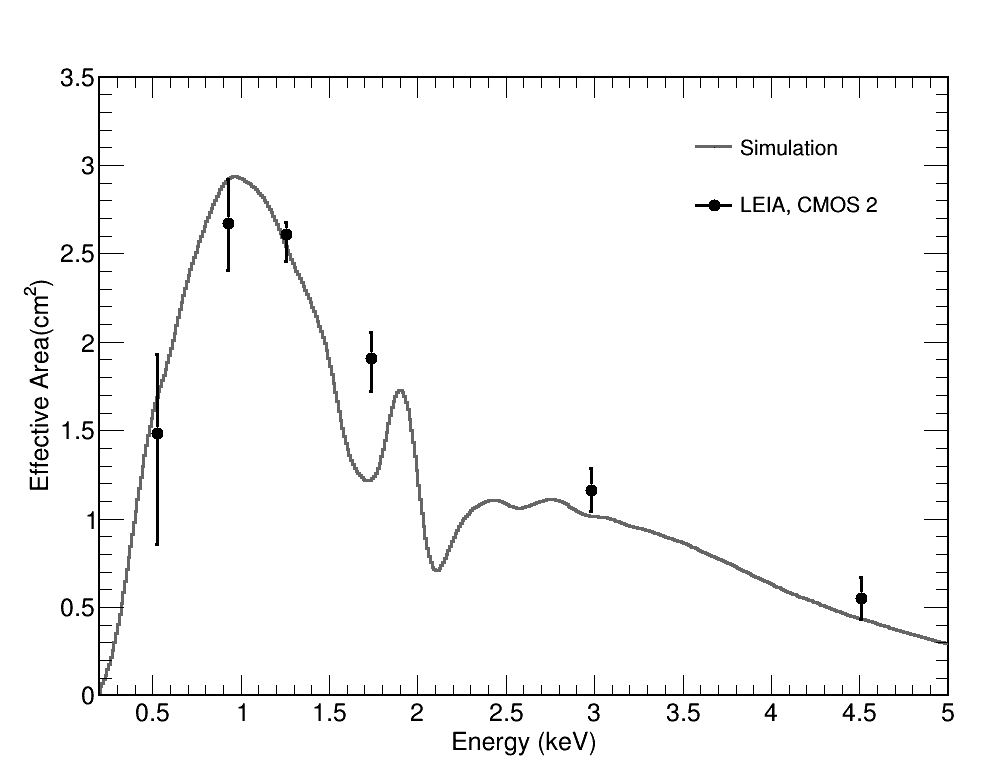}
    \includegraphics[width=0.45\textwidth]{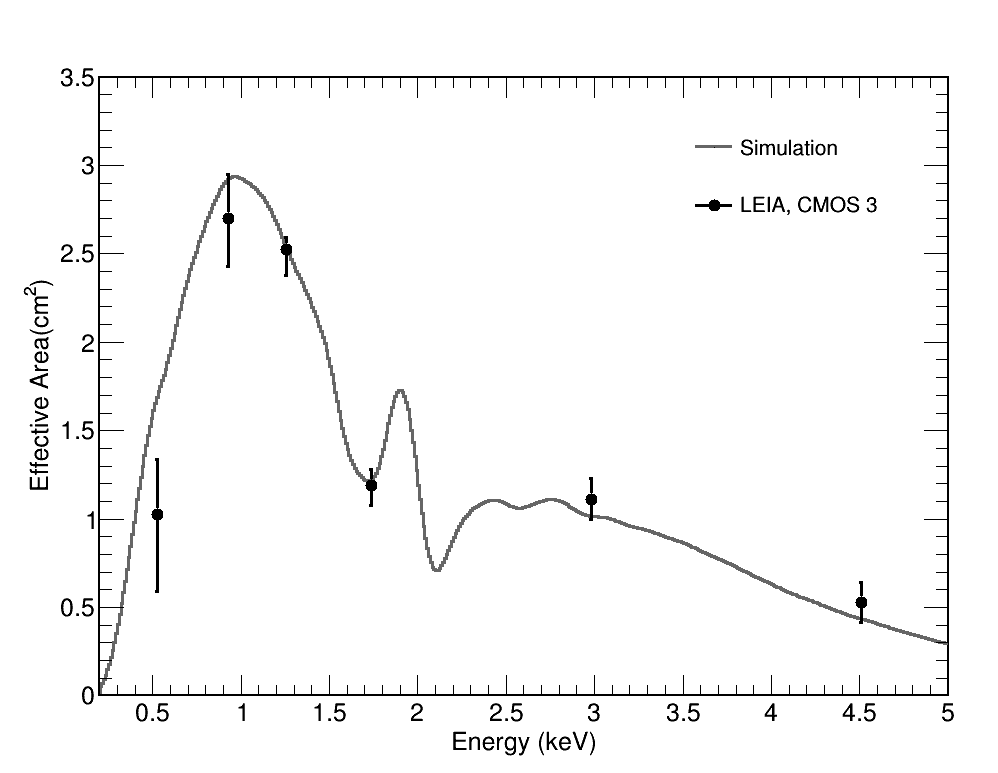}
    \includegraphics[width=0.45\textwidth]{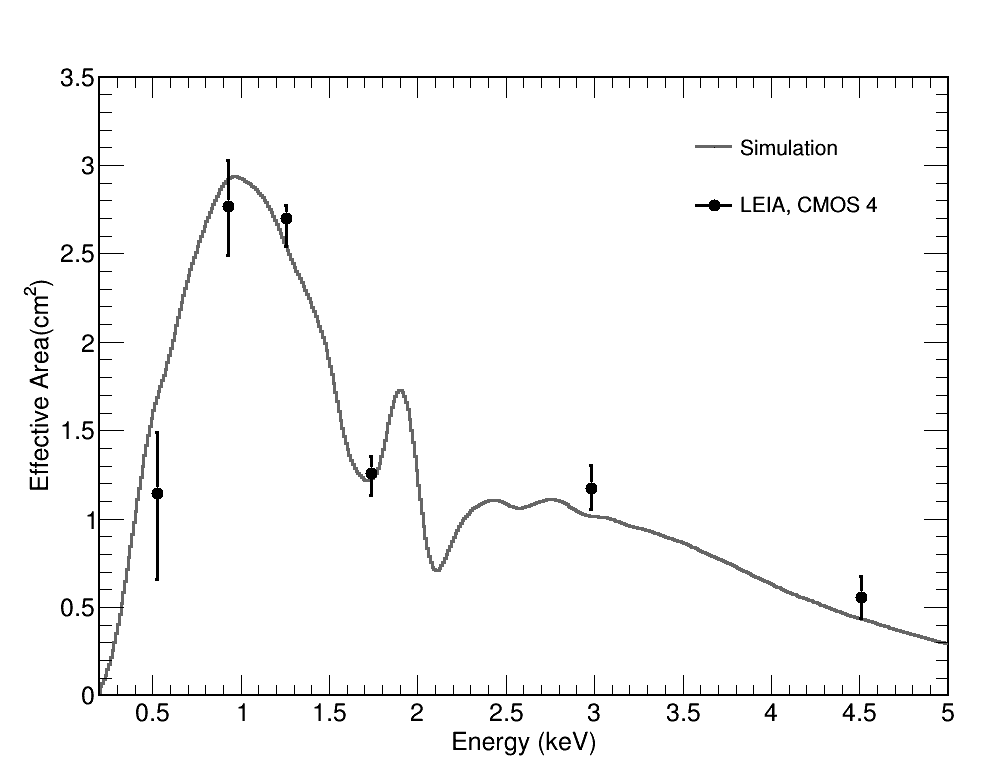}
    \caption{The effective area of the focal spot vs. photon energy, measured in the direction along the center of each CMOS. The black dots are measurements at different test energies. The grey solid line is the simulated model overplotted for comparison.}
    \label{fig:effarea_energy_curve}
\end{figure*}

The effective area of LEIA was measured at various incident angles and energies. A silicon drift detector (SDD), whose quantum efficiency has been carefully calibrated, was used to monitor the beam flux as reference. At a given energy, the effective area ($A_{\rm eff}$) is defined as the factor to convert the photon flux ($\rm cts~s^{-1}~{cm}^{-2}$) of the incident X-ray beam in front of the telescope ($F$) to the measured count rate ($\rm cts~s^{-1}$) by the detectors ($C_{\rm det}$), 
\begin{equation}
\label{eq:effarea}
C_{\rm det} =  A_{\rm eff} \times F 
\end{equation},
where $F$ can be derived from the count rate ($\rm cts~s^{-1}$) of the reference SDD detector ($C_{\rm SDD}$), as
\begin{equation}
\label{eq:fmpo}
    F=\frac{C_{\rm SDD}}{A_{\rm SDD}\times QE_{\rm SDD}}\times G
\end{equation}
where $A_{\rm SDD}=~0.17~{\rm cm^2}$ is the collecting area and $QE_{\rm SDD}$ the quantum efficiency of the SDD. The geometrical correction factor $G$ for the divergent X-ray beam, defined as 
\begin{equation}
\label{eq:geom_coeff}
    G=(d_{\rm source-SDD}/{d_{\rm source-MA}})^2
\end{equation}
where $d_{\rm source-SDD}$ is the distance from source to SDD and $d_{\rm source-MA}$ the distance from source to mirror assembly, is $0.943$ during the experiment. The uncertainty of $A_{\rm eff}$ is contributed both from the statistical errors of the count rates and the systematic error of the quantum efficiency of SDD, and the latter is dominating. 
The $1\sigma$ uncertainty of the measured $A_{\rm eff}$ is calculated by applying the error propagation formalism to Eq. \ref{eq:effarea}.

For each quadrant of the FoV, the effective area was measured for the 121 sampled directions of the full grid at 1.25 keV, 9 directions at 0.53, 1.74, 2.98 and 4.51 keV, and only one direction at 0.93 keV. 
It is found that the measured effective areas are generally consistent with the simulation that was developed in our previous work based on Geant4 \cite{2014SPIE.9144E..4EZ, 2017ExA....43..267Z}.
Fig. \ref{fig:vignetting_figure} shows the distribution of the measured effective area (for the center focal spot only) at 1.25 keV in the 121 directions on the detector plane for CMOS 4, as an example. 
It shows that the effective area ($ 2-3~{\rm cm^2}$) is basically uniform without significant vignetting effects across the FoV except at the edges (due to the incompleteness of the optics beyond the edges).
This result agrees generally with the predicted homogeneity of the lobster eye optics, though a pattern of mild variations is also present to some extent, which is discussed in Section \ref{sec:compare}. 
Fig. \ref{fig:effarea_energy_curve} shows the effective areas (for the center focal spot only) measured at the various line energies; only those in the directions passing through the respective centers of the four detectors are shown, as examples. 
A model prediction obtained from the Monte Carlo simulations \cite{2017ExA....43..267Z} is also over-plotted for comparison. It is found that for most of the energies the measurements match the model in general. The effective area is peaked at $\sim1~{\rm keV}$ ($\sim 3~{\rm cm^2}$), while at $4-5~{\rm keV}$ it decreases to $\sim0.5~{\rm cm^2}$.

\subsection{Performance of the CMOS camera}
\label{sec:cmos_sensor}

\begin{figure*}
\centering
\includegraphics[width=0.48\textwidth]{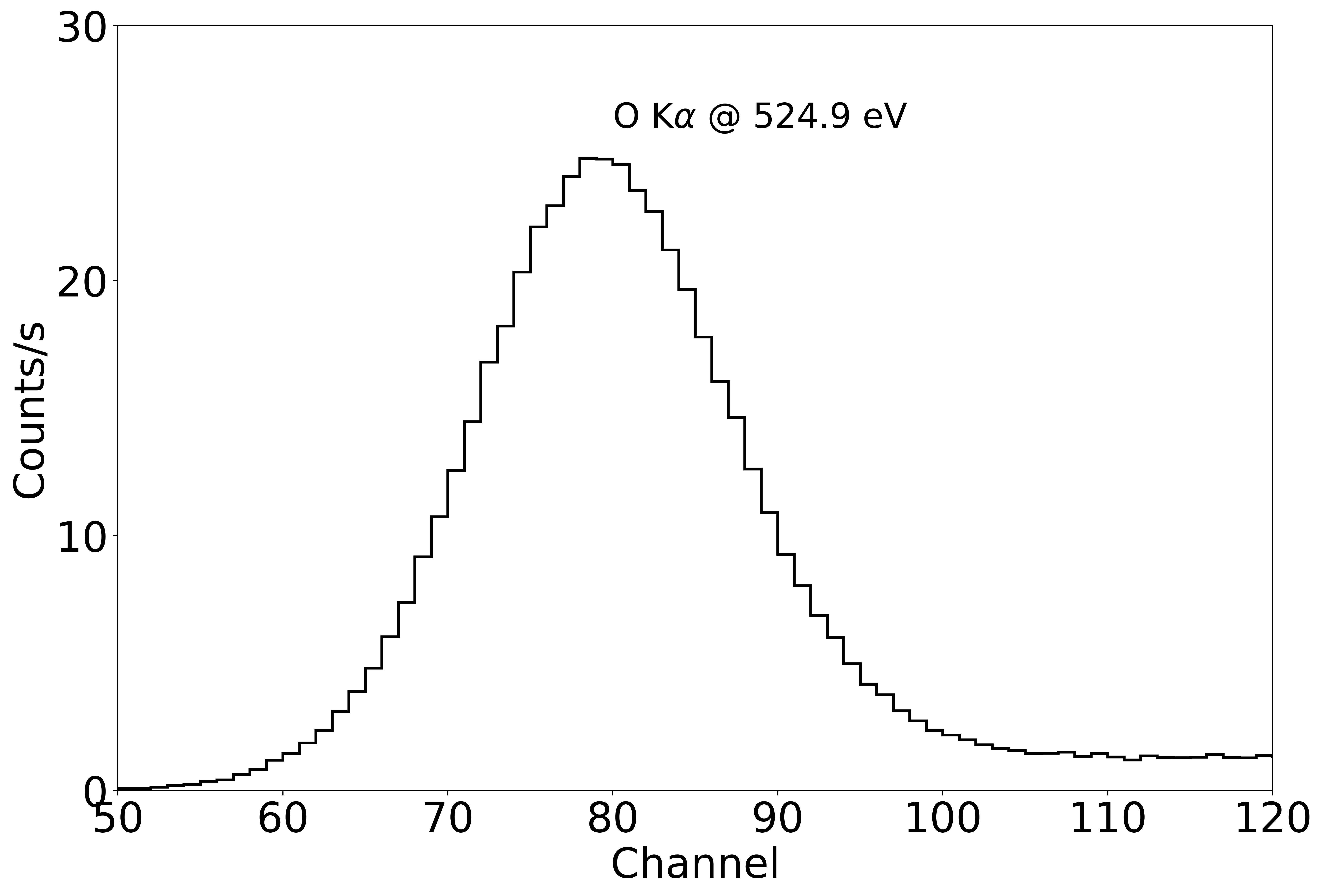}
\includegraphics[width=0.48\textwidth]{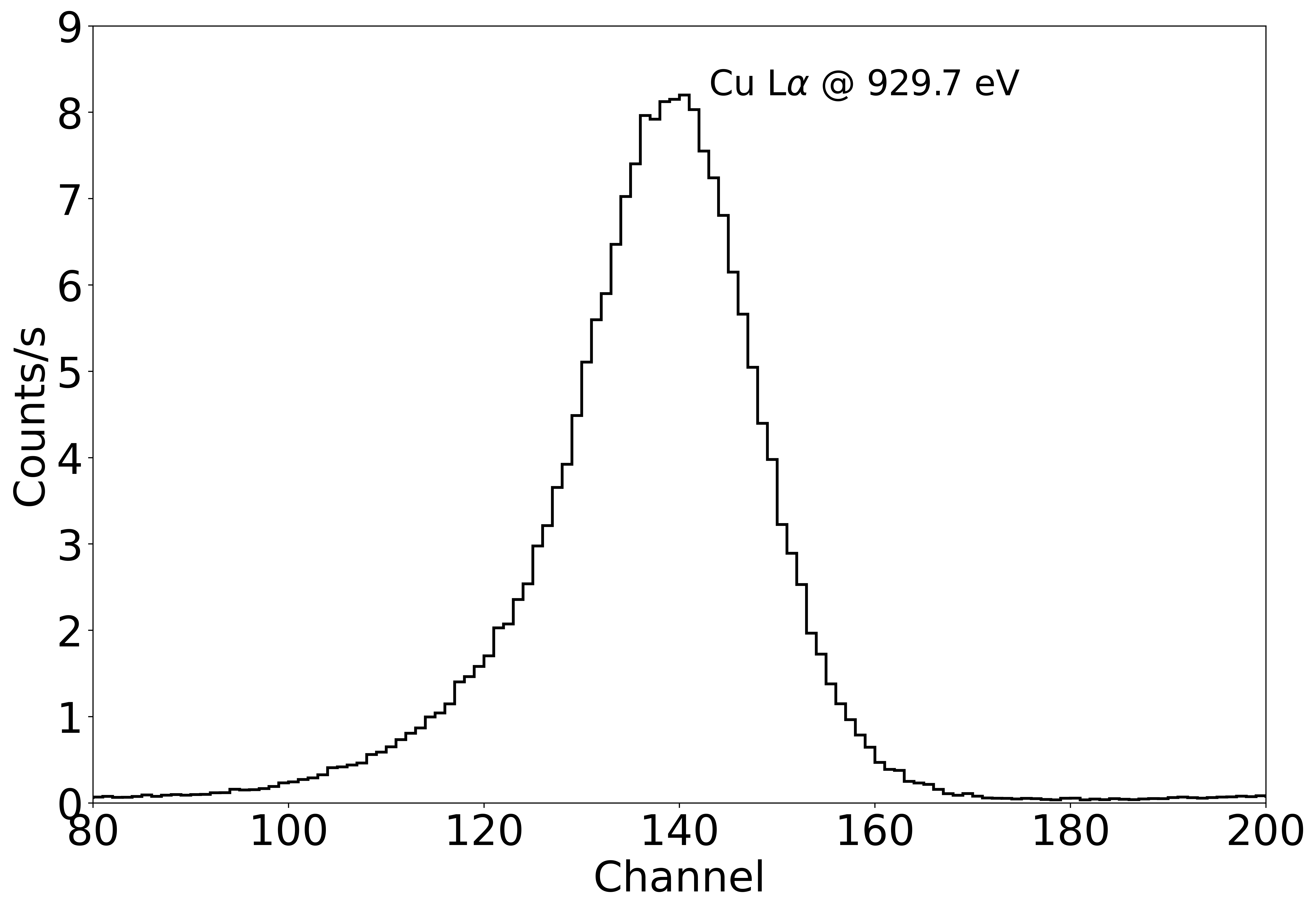}
\includegraphics[width=0.48\textwidth]{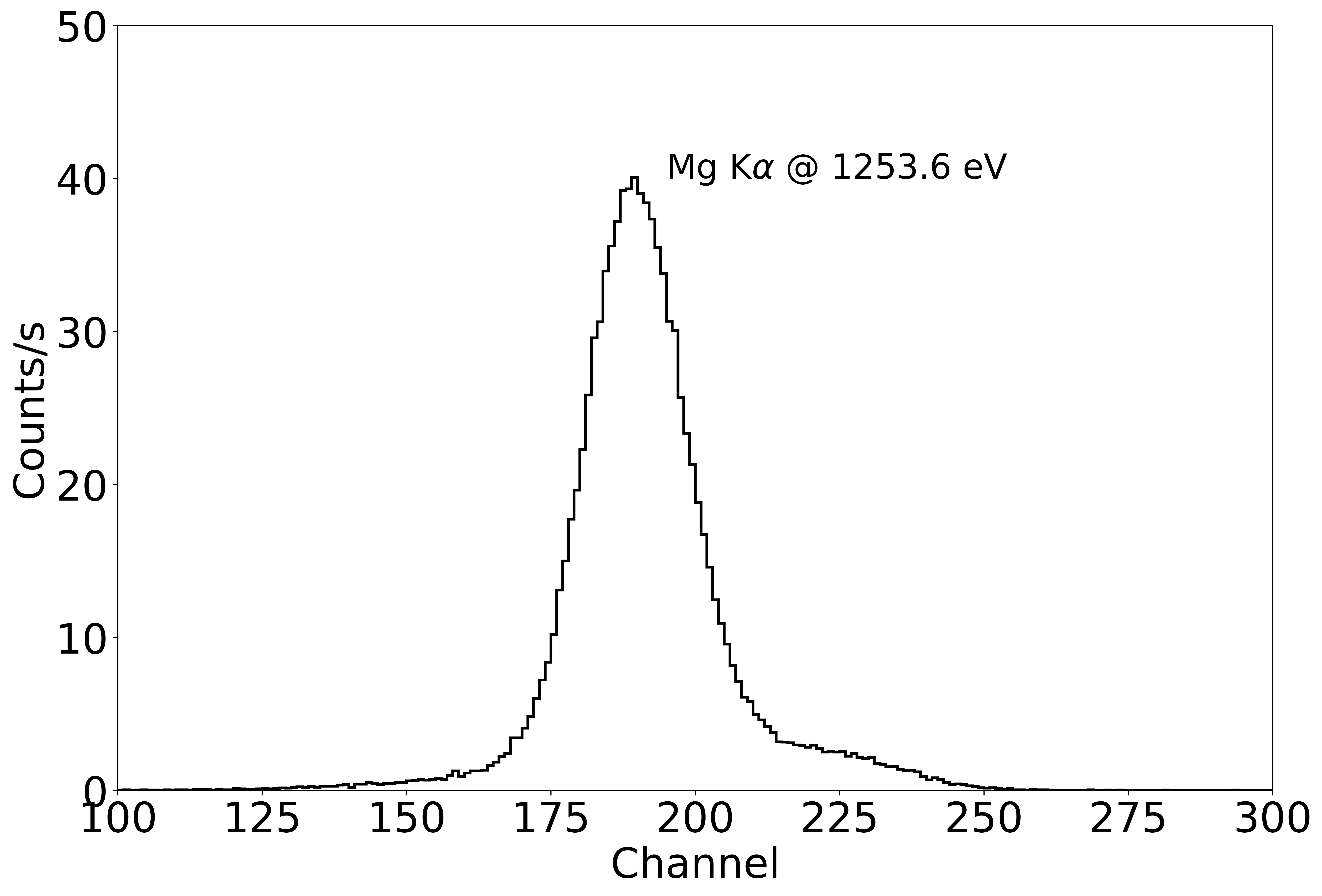}
\includegraphics[width=0.48\textwidth]{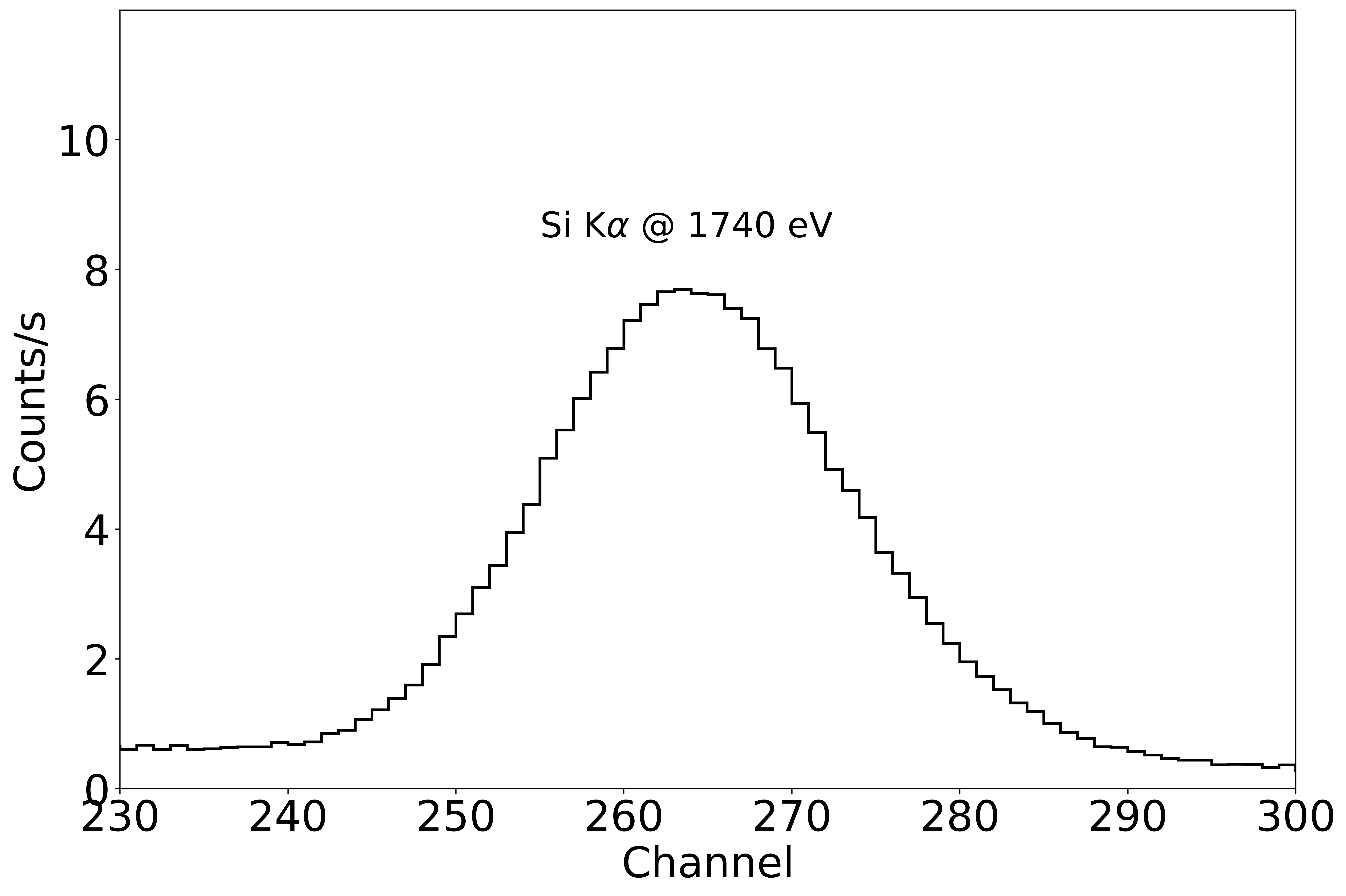}
\includegraphics[width=0.48\textwidth]{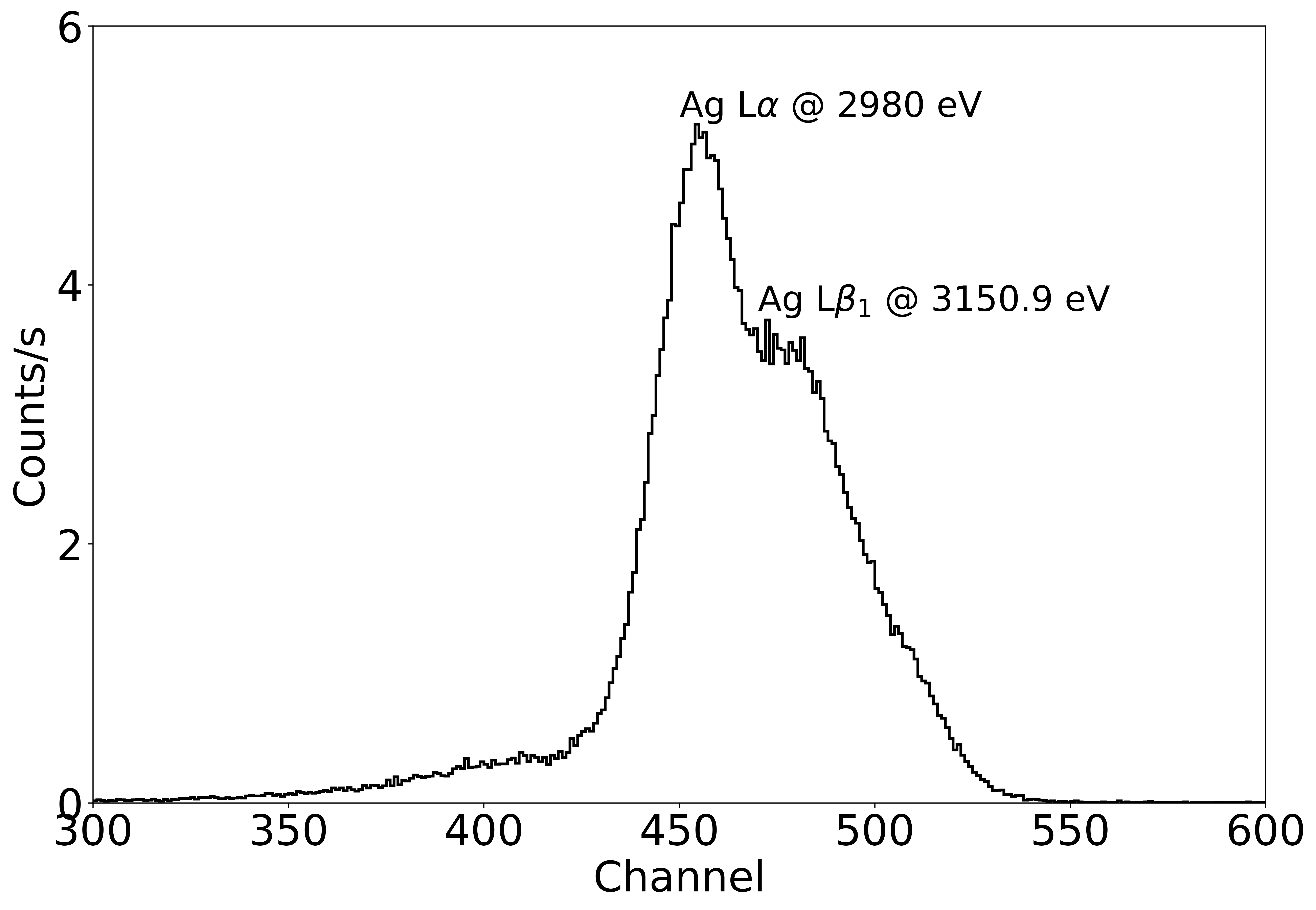}
\includegraphics[width=0.48\textwidth]{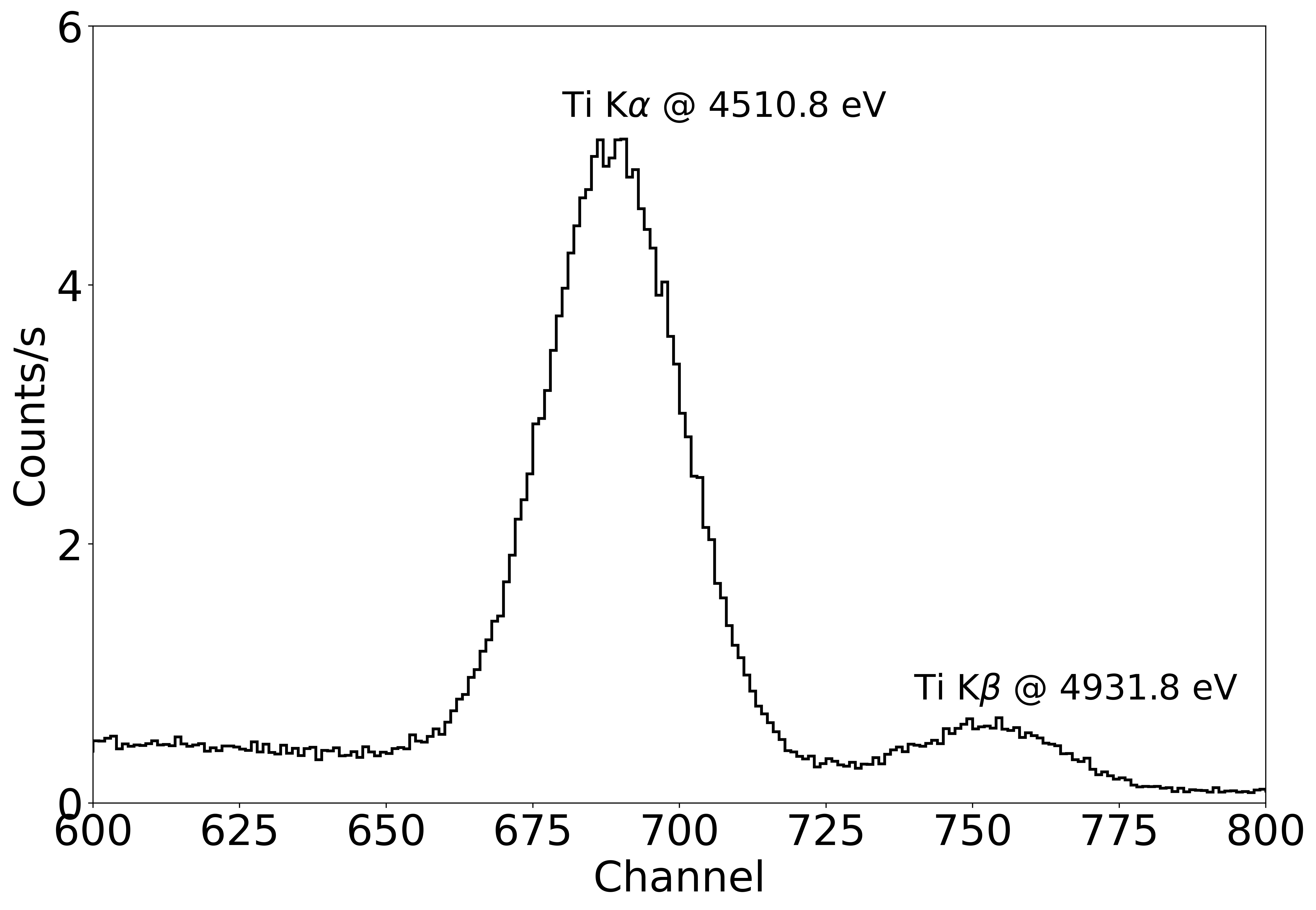}
\caption{\label{fig:cmos_spectra}The measured X-ray spectra for the 6 characteristic emission lines obtained at the center of CMOS 4.}
\end{figure*}

The first tests of the X-ray properties of the four CMOS sensors onboard LEIA were performed individually at NAOC in June 2021 (Ling et al. in preparation). 
After the building of LEIA, the performance of the integrated CMOS camera was calibrated at 100XF, including the energy response, gain, readout noise, bias image and defects. 
As an example, Fig. \ref{fig:cmos_spectra} shows the measured X-ray spectra for the 6 characteristic emission lines obtained at the center of CMOS 4. 
The high-energy tail of the Mg K$\alpha$ line spectrum is mainly due to the contamination by the aluminium filter used in the experiment. 
The asymmetric profile in the Cu L$\alpha$ line at the lower energy end is due to the impurity of the target material. 
A Gaussian fit was used to determine the peak in PHA channels and the width of the line. 

\begin{figure*}
    \centering
    \includegraphics[width=0.75\textwidth]{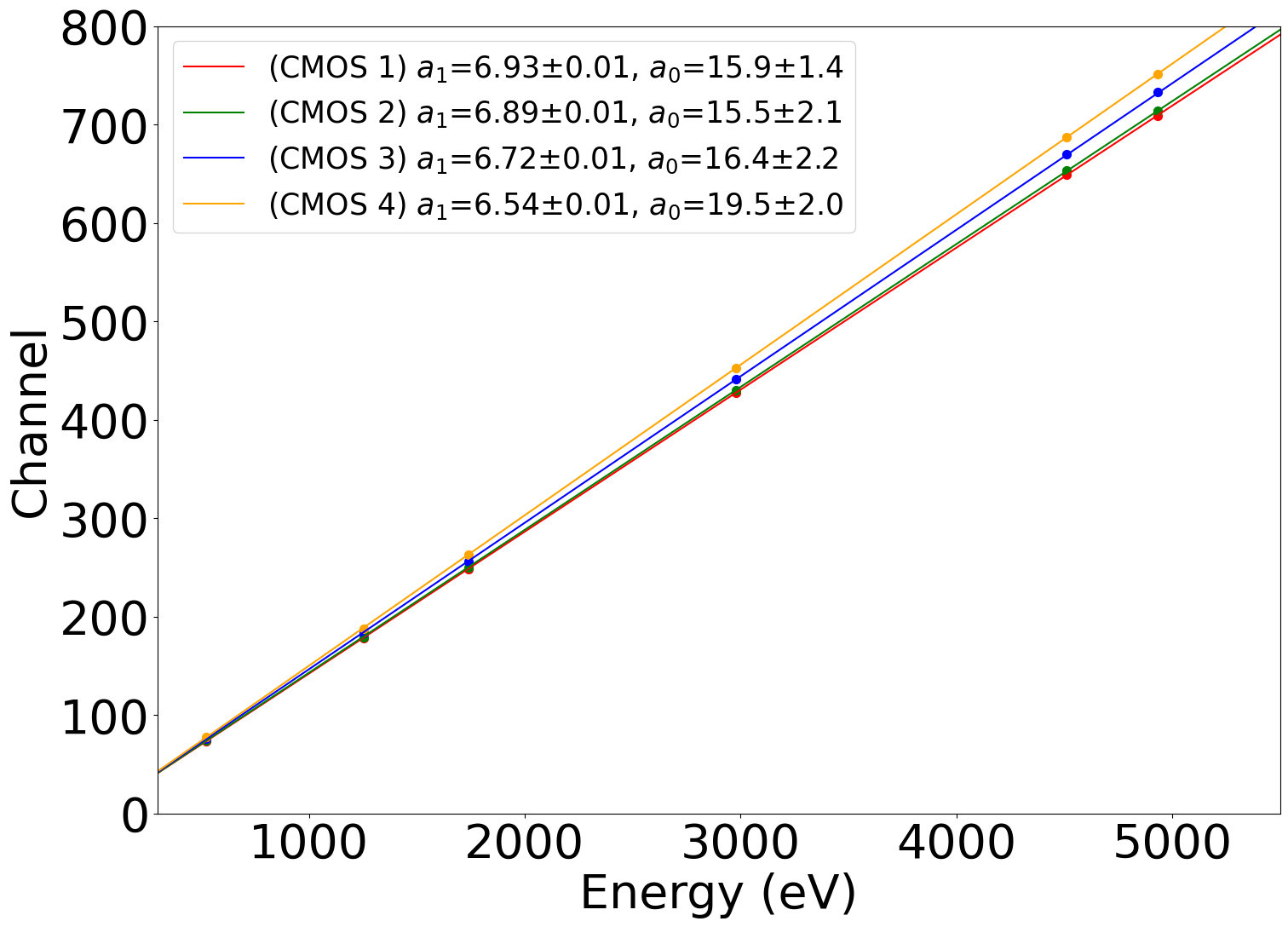}
    \caption{EC relation measured in the direction along the center of each of the CMOS sensors. Different colors refer to different sensors.}
    \label{fig:cmos_gain}
\end{figure*}

We first determined the gain and the energy-channel (EC) relation. 
We choose six emission lines (O K$\alpha$, Mg K$\alpha$, Si K$\alpha$, Ag L$\alpha$, Ti K$\alpha$, and Ti K$\beta$)\footnote{The Cu L$\alpha$ line is excluded due to its asymmetric profile that may introduce uncertainties to the determination of the line energy.} and fit their line energies and the measured peak channels with a linear function
\begin{equation}
    y = a_1\times x + a_0,
\end{equation}
where $x$ is the peak channel in DN and $y$ is the energy of the line in eV.
As an example, Fig. \ref{fig:cmos_gain} shows the EC relation determined in the direction to the center of each of the four CMOS sensors. 
For each sensor, the differences of the gain values measured on the 9 sampled positions are as small as less than 1 percent.
The gain values of the four CMOS sensors are thus in range of $\sim6.5-6.9$ eV/DN.
The energy resolutions (denoted by the FWHM of the fitted Gaussian) also show some differences among different sensors, at a given photon energy. For each of the sensors, the energy resolutions measured at different positions on the CMOS were found to show slight variations. A detailed investigation on the non-uniformity of the energy response of the flight products onboard the EP satellite will be presented elsewhere (Ling et al. in preparation). In general, the energy resolutions for the four sensors are around $120-140$ eV at $1.25$ keV of the Mg K$\alpha$ line and $170-190$ eV at $4.5$ keV of the Ti K$\alpha$ line, as listed in Table \ref{tab:cmos}. 

\begin{table}
\centering
\caption{Characteristics of the four CMOS sensors onboard {\em LEIA}.}
\begin{tabular}{ l | l | l | l | l }
\toprule%
Instrument & CMOS 1 & CMOS 2 & CMOS 3 & CMOS 4 \\ \hline
Readout Noise (e$^{-}$) & 4.5 & 4.6 & 4.4 & 4.7   \\ \hline
Gain & $6.91\pm0.04$ & $6.89\pm0.03$ & $6.70\pm0.03$ & $6.54\pm0.04$ \\ \hline
Energy resolution /eV (0.53 keV) & $117 \pm 3$ & $122 \pm 3$ & $123 \pm 3$ & $116 \pm 2$ \\ \hline
Energy resolution /eV (1.25 keV) & $127 \pm 3$ & $133 \pm 1$ & $133 \pm 7$ & $125 \pm 3$  \\ \hline
Energy resolution /eV (1.74 keV) & $137 \pm 3$ & $143 \pm 2$ & $141 \pm 3$ & $135 \pm 3$ \\ \hline
Energy resolution /eV (4.51 keV) & $175 \pm 2$ & $183 \pm 3$ & $180 \pm 4$ & $174 \pm 2$  \\ \hline
Bad Rows & 1 & 0 & 0 & 0 \\ \hline
Bad Columns & 1 & 0 & 1 & 0 \\ \hline
Bad Pixels & 12 & 4 & 11 & 7 \\ \hline
Bad Regions & 0 & 3 & 2 & 2 \\ 
\botrule
\end{tabular}
\label{tab:cmos}
\end{table}

In Table \ref{tab:cmos}, we summarize the characteristics of the four CMOS sensors onboard LEIA. 
The readout noises of the sensors are about $<5~{\rm e^{-}}$. 
Bad pixels, including single pixels, bad rows and columns, and clusters of bad pixels, were identified and marked. They are excluded from being used onboard. 
The maximum cluster of bad pixels is about 20 by 20 pixels in size, smaller than the typical size of the PSF for LEIA. 

\section{Discussion}
\label{sec:discuss}

\subsection{Comparison with simulation}
\label{sec:compare}

Here we compare the calibration results with the simulations carried out for the EP WXT module \cite{2014SPIE.9144E..4EZ,2017ExA....43..267Z,2021OptCo.48326656L,2022PASP..134k5002L}. We mainly focus on the properties of the PSF and effective area, which were both measured at various energies of characteristic X-ray emission lines and in different incident directions.

For the measurements of the PSF at 1.25 keV (using the Mg K$\alpha$ line), a total of $22\times22$ directions (484 points on the detectors) were sampled across the FoV (see Fig. \ref{fig:psfscan_mgk}). 
Apart from the regions subject to the misalignment issue (see Section \ref{sec:psf_misalignment} for discussion), the characteristic cruciform shapes of the PSFs show reasonably well consistency among most of the sampled directions, as predicted by the lobster eye optics.
The width of the central focal spots falls mostly in the range of 4--7 arcmin (the long axis and area-equivalent radius, see Fig. \ref{fig:psf_fitting_cdf}).
As can be seen, the PSFs shows mild non-uniformity across the FoV in both shape and size, which deviates from the prediction for idealised, perfect lobster-eye optics. 
This is mainly caused by the inherent imperfectness introduced in the manufacturing and mounting of the MPO optics, as well as the mismatch between the flat detector plane and the spherical focal surface.

The elliptical shape of the focal spot can be well reproduced by simulations when the distortions of the micro-square pore channels are taken into account \cite{2021OptCo.48326656L, 2022PASP..134k5002L}. It is found that low energy photons spread over a larger area than those at high energies on the detector (Fig. \ref{fig:psf_dif_en}), which is also well consistent with Monte Carlo simulations \cite{2017ExA....43..267Z}. This is because the low energy photons have a larger critical angle of grazing incidence thus they can be reflected at larger angles than higher energy ones.

Theoretically, lobster eye optics are free from vignetting except near the edge of the FoV, as they are spherically symmetric in essentially all directions.
In this sense, the effective areas measured in most of the sampled directions should be similar, as shown in Fig. \ref{fig:vignetting_figure}.
The mild non-uniformity of the effective area within the FoV (except for the edge region) is mainly due to the obscuration of  the incident X-ray photons  by the mounting frame between the individual MPO plates, resulting in a smaller effective area in these directions than the nominal $\sim3~{\rm cm^2}$ elsewhere. Such a distribution pattern of the effective area can be well reproduced by simulations and will be presented in a separate paper (Zhao et al. in preparation). 

The dependence of the effective area on incident photon energy was investigated at the characteristic line energies of O K$\alpha$, Cu L$\alpha$, Mg K$\alpha$, Si K$\alpha$, Ag L$\alpha$, and Ti K$\alpha$. It is found that at most of the energies the measurements generally agree with models generated by simulations based on Geant4 (see Fig. \ref{fig:effarea_energy_curve}).
We note that the relatively large discrepancy at the lowest measured energy of O K$\alpha$ line is likely due to the large uncertainties of the quantum efficiency of the SDD detector (as denoted by the large error bars). 
On the other hand, the larger effective areas measured at 0.53 keV (O K$\alpha$) and 1.74 keV (Si K$\alpha$) for the FoV quadrant subtended by CMOS 2 compared to other quadrants is likely a result of underestimation of the beam strength. 
Specifically, during when the quadrant was measured with the beam light generated by the SiO$_2$ source, there was a significant decrease of the SDD count rate by as large as $\sim40$ percent, while the count rate measured simultaneously by CMOS 2 was generally stable (indicating no significant variations in the beam light). 

To summarize, the distributions of the PSF and effective areas across the whole FoV are largely uniform, although they also show mild non-uniformity to some extent. 
The deviations from the prediction of idealized, perfect lobster-eye optics can be understood to be caused by the imperfect shapes and alignment of the micro-pores as well as the obscuration of photons by the supporting frames, which can be well reproduced by Monte Carlo simulations. 
The dependence of both the PSF and effective area on the energy of incident photons is also in general agreement with the simulation results.

\subsection{Possible effects on the calibration results}
\label{sec:factors}

Here we discuss several possible effects in the experiment that may affect the calibration results, namely the defocus effect due to the finite distance of the source, the stability of the X-ray source, and the operating temperatures of the instrument and assemblies. 

The focal detector module is mounted at a position optimized to best match the focal sphere of the MA (focal length $375~{\rm mm}$) assuming an infinite distance of the source.
In the on-ground calibration experiment, however, the point source is located at a finite distance. This de-focus effect would inevitably affect the measured instrumental performance. 
Specifically, for a source distance of $\sim100$ meters as the case for our experiments, the image distance is $\sim1.4$ mm shorter of the designed focal length. 
Such a defocus effect would, according to our previous simulations \cite{2017ExA....43..267Z}, result in only a mild broadening of the focal spot PSF by $\sim0.2$ arcmin in FWHM and a negligible decrease of the effective area by less than $\sim2$ percent in $0.5-4$ keV.
These deviations are smaller than or comparable to the measurement errors.  
We thus conclude that the defocus in our experiment at the 100XF has negligible effects on the measured instrumental properties, which can be served as a reasonable approximation to the in-flight performance for infinite sources. 
A more detailed investigation in combination with simulation results will be presented elsewhere (Zhao et al. in preparation).

To obtain the effective area, the strength of the X-ray beam light was measured by using a reference SDD detector mounted beside the MA at a monitoring cadence of once every $5-10$ minutes (the accumulation time for the SDD detector varies slightly for different targets lines). 
Overall, the beam light was found to be largely stable during the experiment, with mild short-term fluctuations (typically, at a level of a few percent) and long-term variations on timescales of hours, however. 
For instance, during the experiment using the Mg source, there was a gradual increase of the beam strength by $\sim1-2$ percent within a day. 
We argue that the variations in the beam light would not introduce notable systematics to the measured properties.
For the PSF measurement, a sufficiently long exposure time was adopted to achieve high photon statistics in the measured image. 
For the effective area measurement, the beam strength measured simultaneously with the reference SDD detector was adopted to calculate the effective area. 

The performances of the instrument, particularly those related to the CMOS sensors, may be affected by the temperature of the device and its variation during the experiment. 
During all the calibration experiments, the MA and detector modules were maintained at their nominal in-orbit working temperatures of $\sim20$ and $-30$ degrees, respectively, 
and no significant variations were found during the entire campaign.
It is worth noting that, a previous thermal test performed on a separate WXT qualification model at the 100XF shows no noticeable variations in the optical performance as the whole MA is heated uniformly all the way up to $25$ degrees. 
Furthermore, in a thermal experiment on another qualification model of the MA carried out at the University of Leicester  as a collaborative work \cite{2020charly_wxt_LU}, it was found that the imaging quality of the MA remains almost unchanged with a thermal gradient of $\sim9~{\rm ^{o}C}$ across the optic (private communication, Feldman Charlotte).
The performance of the CMOS sensors onboard LEIA operated at various temperatures will be presented elsewhere (Ling et al. in preparation). 

To summarize, the calibration results presented above are basically little or not affected by the possible effects considered, mainly de-focus, the stability of the X-ray source and the operating temperatures of the instrument. 

\subsection{Misalignment of the PSF}
\label{sec:psf_misalignment}

\begin{figure*}
    \centering
    \includegraphics[width=\textwidth]{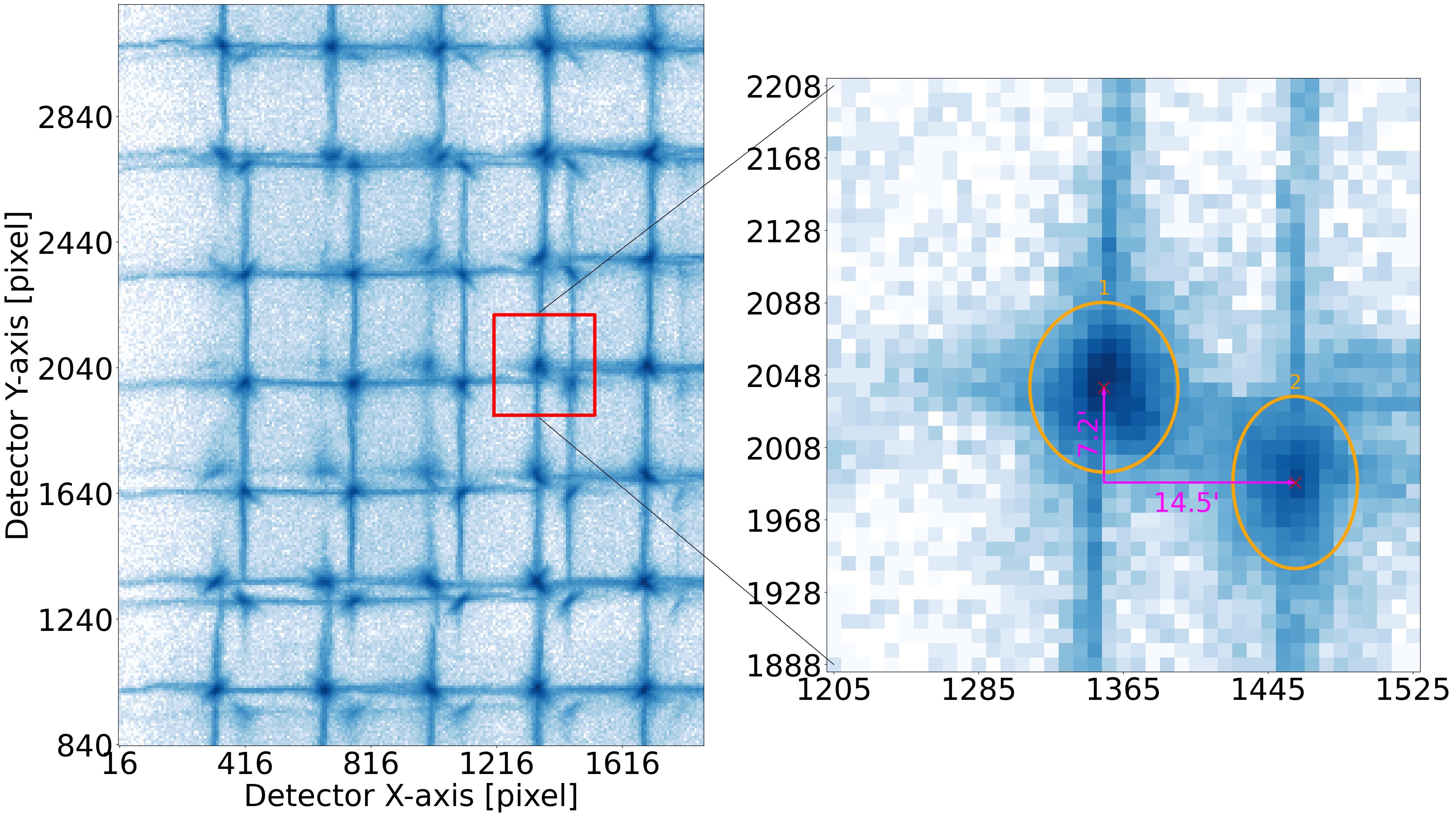}
    \caption{({\em Left}): A zoom-in image for the Region 3 on CMOS 2 (upper-right panel of Fig. \ref{fig:psfscan_mgk}). ({\em Right}): A further zoom-in image for a $320\times320$ pixel region where the misalignment results in a splitting of the PSF into double focal spots and arms. The angular distances for the two detached focal spots are $\sim14.5'$ and $\sim7.2'$ in X- and Y- directions, respectively.}
    \label{fig:focalspot_misalignment_zoomin}
\end{figure*}

As shown in Section \ref{sec:psf_dif_direction}, the PSFs in three confined regions within the FoV of LEIA, two on CMOS 1 (Region 1 and Region 2 in the upper left panel of Fig. \ref{fig:psfscan_mgk}) and one on CMOS 2 (Region 3 in the upper right panel of Fig. \ref{fig:psfscan_mgk}), are misaligned with respect to the PSFs in the rest of the FoV.
The total area of misalignment takes up about 9\% of the entire FoV.
A zoom-in image of Region 3 is shown in Fig. \ref{fig:focalspot_misalignment_zoomin}. 
An apparent feature is the splitting of the PSF into double or triple focal spots and arms, which is particularly discernible near the boundaries of these regions where the misaligned spots contributed by two or three adjacent MPO plates are comparable in strength. 
For the region shown in Fig. \ref{fig:focalspot_misalignment_zoomin}, the offsets of the two focal spots are around $14.5$ arcmin along the X-axis and $7.2$ arcmin along the Y-axis, respectively. The range of the detectors and the level of misalignment are listed in Table \ref{table:psfmisalign}. In general, the offsets of the PSF are around $10-20$ arcmin, significantly larger than the systematics of source positioning of $\sim 1-2$ arcmin \cite{2023LingZXRAA}. 
Consequently, the analysis of the data collected within these regions should be treated with caution. An algorithm for the PSF reconstruction and optimization to address this issue is now in progress. 

The causes of the PSF misalignment are twofold. For Region 1 and Region 3, one of the MPO plates responsible for imaging in this region was inaccurately mounted with a displacement with respect to the other confocal MPO plates, so that its focus is offset to the common focal point.  
For region 2, the slight offset of the focal spot, which has the least extent among the three misalignment regions, is due to the considerably poorer imaging quality of the MPO plates than the rest MPO of the MA of LEIA. 

\begin{table}
\caption{The range of the detectors with PSF misalignment and the level of misalignment.}
\centering
\begin{tabular}{c|c|c|c|c|c}
\hline
Region & CMOS & DETX range & DETY range & Offset in X & Offset in Y\\ \hline
1 & 1 & (0, $\sim1800$) & (0, $\sim1800$) & $11.2^{'} \pm1.3^{'}$ & $7.1^{'} \pm2.7^{'}$ \\ \hline
2 & 1 & ($\sim1200$, $\sim2800$) & ($\sim1800$, $\sim3200$) & $1.1^{'}\pm0.9^{'}$ & $8.8^{'}\pm1.0^{'}$ \\ \hline
3 & 2 & (0, $\sim1800$) & ($\sim800$, $\sim3200$) & $12.9^{'}\pm1.6^{'}$ & $7.3^{'}\pm2.7^{'}$ \\ \hline
\end{tabular}
\label{table:psfmisalign}
\end{table}

\section{Summary}
\label{sec:summary}

In this paper, we report on results of the end-to-end calibration of the LEIA instrument carried out at the 100-m X-ray Test Facility at IHEP, CAS in November, 2021. 
This paper focuses mainly on the properties of the PSF, effective area and energy response, of which the measurements were made in a wide range of incident directions and at various characteristic X-ray line energies. 
Specifically, for each of the FoV quadrants, the performances were measured for a sampling grid of 11 $\times$ 11 points on the detector, corresponding to 121 incident directions across the FoV at the line energy of Mg K$\alpha$. For the line energies of O K$\alpha$, Si K$\alpha$, Ag L$\alpha$, and Ti K$\alpha$, the measurements were carried out for a sub-grid of 3 $\times$ 3 points corresponding to 9 directions within the FoV. For the energy of Cu L$\alpha$ line only the direction passing through the CMOS chip center was sampled.

The characteristic cruciform shapes of the PSFs are found to be largely uniform and consistent among most of the sampled directions across the FoV, which are largely in agreement with the prediction of lobster-eye optics.
The elliptical shape of the PSF focal spot, as well as the mild non-uniformity of the PSF in shape and size are mainly caused by the imperfect shapes and alignment of the micro-pores, and can be well reproduced by Monte Carlo simulations.
The spatial resolution defined as the FWHM measured in the direction of long axis of the focal spot, ranges from 4 to 8 arcmin with a median of 5.7 arcmin. 
The decreasing area of photon spreading onto the detector with increasing photon energy also agrees with theoretical predictions and simulation results.

The roughly uniform distribution of the effective areas of the center focal spot ($2-3~{\rm cm^2}$ at $1.25~{\rm keV}$) across the FoV is in general agreement with the theoretical prediction of lobster eye optics that there is almost no vignetting effect across the FoV (except near the edges). 
The mild variations of the effective area arise mainly from the obscuration of the incident X-ray photons by the mounting frame between the individual MPO plates, and can be well reproduced by simulations. 
The measured effective areas at different energies also agree largely with the models generated by our previous simulations. 

The energy response of the CMOS detectors is also investigated. For the four CMOS sensors, the gains are in the range of $6.5-6.9$ eV/DN, and the energy resolutions are in the range of $\sim120-140$ eV at $1.25$ keV and $\sim170-190$ eV at $4.5$ keV. 
For each individual sensor, there is only a mild non-uniformity of the energy response for different regions on the sensors that were sampled.
The readout noises of the sensors are $<5~{\rm e^{-}}$. Bad pixels have been identified from the analysis of the bias maps and will be excluded from being used in future data taking.

Several effects that could potentially affect the accuracy of the calibration, namely the defocus effect, the stability of the X-ray source and the operating temperatures of the instrument and assemblies, are discussed. They are found to have little or no effect on the measured properties of the instrument.
The PSF misalignment in three confined regions, which affects a total of 9\% of the entire FoV, is mainly caused by inaccurate mounting of one of the MPO plates, respectively, as well as the poor imaging quality of the few corresponding MPO plates. 

The on-ground calibration results have already been ingested into the calibration database (1st version) and applied to the analysis of the scientific data acquired by LEIA. 
LEIA has been operating in orbit for more than a year so far. 
In-flight calibration observations of a few bright X-ray sources (e.g. Sco X-1, Crab nebula) show no significant variations in the imaging quality and effective area compared to the measurements in the on-ground calibration. 
The results of the in-flight calibration will be presented elsewhere (Cheng et al. in preparation). 
The calibration work of LEIA also paves the way for the calibration of the flight model of the Wide-field X-ray Telescope of the Einstein Probe mission. 

\ 

\

\bmhead{Acknowledgments}
We thank all the members of the EP team, the EP consortium, and the SATech-01 team. The CAS team acknowledge contribution from ESA for calibration of the mirror assembly and tests of part of the devices. The Leicester and MPE teams acknowledge funding by ESA. 

\bmhead{Author’s Contribution}
H.-Q.C. and Z.-X.L. led the paper writing. C.Z. led the calibration experiment design and data analysis. W.-M.Y.contributed to the improvement of the manuscript. Z.-X.L., C.Z., W.-X.W., Y.-F.D, D.-H.Z., Z.-Q.J., X.-J.S., S.-L.S., Y.-F.C., Z.-W.C., W.F., Y.-X.H., J.-F.L., Z.-D.L., X.-H.M., Y.-L.X., A.-L.Y., Q.Z. and W.-M.Y. contributed to the design and development the instrument. H.-Q.C., Z.-X.L., C.Z.,D.-H.Z., Y.-F.D., Z.-Q.J., X.-J.S., Y.-F.C., Z.-W.C., Y.-X.H., J.-F.L., Z.-D.L., X.-H.M., Y.-L.X. participated in the calibration experiment. H.-Q.C., Z.-X.L., C.Z., Y.L., H.-W.P., W.-X.W., D.-H.Z. contributed to the analysis of calibration data. Y.-S.W. is responsible for the construction of the test facility. X.-T.Y. and Z.-J.Z. led the experimental setup and data acquisition. All authors read and approved the final manuscript.

\bmhead{Funding}
This work is supported by the Einstein Probe project, a mission in the Strategic Priority Program on Space Science of CAS (grant Nos. XDA15310000, XDA15052100), and the National Science Foundation of China (grant no. 12173055, 12173057, 12203071). 

\bmhead{Availability of data and material}
The datasets used and analysed during the current study are available from the first corresponding author on reasonable request.

\section*{Declarations}

\bmhead{Conflicts of interest}
The authors declare that they have no conflict of interest.


\bibliography{leia_calib}


\begin{thebibliography}{22}
\ifx \bisbn   \undefined \def \bisbn  #1{ISBN #1}\fi
\ifx \binits  \undefined \def \binits#1{#1}\fi
\ifx \bauthor  \undefined \def \bauthor#1{#1}\fi
\ifx \batitle  \undefined \def \batitle#1{#1}\fi
\ifx \bjtitle  \undefined \def \bjtitle#1{#1}\fi
\ifx \bvolume  \undefined \def \bvolume#1{\textbf{#1}}\fi
\ifx \byear  \undefined \def \byear#1{#1}\fi
\ifx \bissue  \undefined \def \bissue#1{#1}\fi
\ifx \bfpage  \undefined \def \bfpage#1{#1}\fi
\ifx \blpage  \undefined \def \blpage #1{#1}\fi
\ifx \burl  \undefined \def \burl#1{\textsf{#1}}\fi
\ifx \doiurl  \undefined \def \doiurl#1{\url{https://doi.org/#1}}\fi
\ifx \betal  \undefined \def \betal{\textit{et al.}}\fi
\ifx \binstitute  \undefined \def \binstitute#1{#1}\fi
\ifx \binstitutionaled  \undefined \def \binstitutionaled#1{#1}\fi
\ifx \bctitle  \undefined \def \bctitle#1{#1}\fi
\ifx \beditor  \undefined \def \beditor#1{#1}\fi
\ifx \bpublisher  \undefined \def \bpublisher#1{#1}\fi
\ifx \bbtitle  \undefined \def \bbtitle#1{#1}\fi
\ifx \bedition  \undefined \def \bedition#1{#1}\fi
\ifx \bseriesno  \undefined \def \bseriesno#1{#1}\fi
\ifx \blocation  \undefined \def \blocation#1{#1}\fi
\ifx \bsertitle  \undefined \def \bsertitle#1{#1}\fi
\ifx \bsnm \undefined \def \bsnm#1{#1}\fi
\ifx \bsuffix \undefined \def \bsuffix#1{#1}\fi
\ifx \bparticle \undefined \def \bparticle#1{#1}\fi
\ifx \barticle \undefined \def \barticle#1{#1}\fi
\bibcommenthead
\ifx \bconfdate \undefined \def \bconfdate #1{#1}\fi
\ifx \botherref \undefined \def \botherref #1{#1}\fi
\ifx \url \undefined \def \url#1{\textsf{#1}}\fi
\ifx \bchapter \undefined \def \bchapter#1{#1}\fi
\ifx \bbook \undefined \def \bbook#1{#1}\fi
\ifx \bcomment \undefined \def \bcomment#1{#1}\fi
\ifx \oauthor \undefined \def \oauthor#1{#1}\fi
\ifx \citeauthoryear \undefined \def \citeauthoryear#1{#1}\fi
\ifx \endbibitem  \undefined \def \endbibitem {}\fi
\ifx \bconflocation  \undefined \def \bconflocation#1{#1}\fi
\ifx \arxivurl  \undefined \def \arxivurl#1{\textsf{#1}}\fi
\csname PreBibitemsHook\endcsname

\bibitem[\protect\citeauthoryear{{Angel}}{1979}]{1979ApJ...233..364A}
\begin{barticle}
\bauthor{\bsnm{{Angel}}, \binits{J.R.P.}}:
\batitle{{Lobster eyes as X-ray telescopes.}}
\bjtitle{\apj}
\bvolume{233},
\bfpage{364}--\blpage{373}
(\byear{1979})
\doiurl{10.1086/157397}
\end{barticle}
\endbibitem

\bibitem[\protect\citeauthoryear{{Fraser} et~al.}{1992}]{1992SPIE.1546...41F}
\begin{bchapter}
\bauthor{\bsnm{{Fraser}}, \binits{G.W.}},
\bauthor{\bsnm{{Lees}}, \binits{J.E.}},
\bauthor{\bsnm{{Pearson}}, \binits{J.F.}},
\bauthor{\bsnm{{Sims}}, \binits{M.R.}},
\bauthor{\bsnm{{Roxburgh}}, \binits{K.}}:
\bctitle{{X-ray focusing using microchannel plates}}.
In: \beditor{\bsnm{{Hoover}}, \binits{R.B.}} (ed.)
\bbtitle{Multilayer and Grazing Incidence X-Ray/EUV Optics}.
\bsertitle{Society of Photo-Optical Instrumentation Engineers (SPIE) Conference
  Series},
vol. \bseriesno{1546},
pp. \bfpage{41}--\blpage{52}
(\byear{1992}).
\doiurl{10.1117/12.51224}
\end{bchapter}
\endbibitem

\bibitem[\protect\citeauthoryear{{Willingale}
  et~al.}{1998}]{1998ExA.....8..281W}
\begin{barticle}
\bauthor{\bsnm{{Willingale}}, \binits{R.}},
\bauthor{\bsnm{{Fraser}}, \binits{G.W.}},
\bauthor{\bsnm{{Brunton}}, \binits{A.N.}},
\bauthor{\bsnm{{Martin}}, \binits{A.P.}}:
\batitle{{Hard X-ray imaging with microchannel plate optics}}.
\bjtitle{Experimental Astronomy}
\bvolume{8}(\bissue{4}),
\bfpage{281}--\blpage{296}
(\byear{1998})
\end{barticle}
\endbibitem

\bibitem[\protect\citeauthoryear{{Fraser} et~al.}{2002}]{2002SPIE.4497..115F}
\begin{bchapter}
\bauthor{\bsnm{{Fraser}}, \binits{G.W.}},
\bauthor{\bsnm{{Brunton}}, \binits{A.N.}},
\bauthor{\bsnm{{Bannister}}, \binits{N.P.}},
\bauthor{\bsnm{{Pearson}}, \binits{J.F.}},
\bauthor{\bsnm{{Ward}}, \binits{M.}},
\bauthor{\bsnm{{Stevenson}}, \binits{T.J.}},
\bauthor{\bsnm{{Watson}}, \binits{D.J.}},
\bauthor{\bsnm{{Warwick}}, \binits{B.}},
\bauthor{\bsnm{{Whitehead}}, \binits{S.}},
\bauthor{\bsnm{{O'Brian}}, \binits{P.}},
\bauthor{\bsnm{{White}}, \binits{N.}},
\bauthor{\bsnm{{Jahoda}}, \binits{K.}},
\bauthor{\bsnm{{Black}}, \binits{K.}},
\bauthor{\bsnm{{Hunter}}, \binits{S.D.}},
\bauthor{\bsnm{{Deines-Jones}}, \binits{P.}},
\bauthor{\bsnm{{Priedhorsky}}, \binits{W.C.}},
\bauthor{\bsnm{{Brumby}}, \binits{S.P.}},
\bauthor{\bsnm{{Borozdin}}, \binits{K.N.}},
\bauthor{\bsnm{{Vestrand}}, \binits{T.}},
\bauthor{\bsnm{{Fabian}}, \binits{A.C.}},
\bauthor{\bsnm{{Nugent}}, \binits{K.A.}},
\bauthor{\bsnm{{Peele}}, \binits{A.G.}},
\bauthor{\bsnm{{Irving}}, \binits{T.H.}},
\bauthor{\bsnm{{Price}}, \binits{S.}},
\bauthor{\bsnm{{Eckersley}}, \binits{S.}},
\bauthor{\bsnm{{Renouf}}, \binits{I.}},
\bauthor{\bsnm{{Smith}}, \binits{M.}},
\bauthor{\bsnm{{Parmar}}, \binits{A.N.}},
\bauthor{\bsnm{{McHardy}}, \binits{I.M.}},
\bauthor{\bsnm{{Uttley}}, \binits{P.}},
\bauthor{\bsnm{{Lawrence}}, \binits{A.}}:
\bctitle{{LOBSTER-ISS: an imaging x-ray all-sky monitor for the International
  Space Station}}.
In: \beditor{\bsnm{{Flanagan}}, \binits{K.A.}},
\beditor{\bsnm{{Siegmund}}, \binits{O.H.W.}} (eds.)
\bbtitle{X-Ray and Gamma-Ray Instrumentation for Astronomy XII}.
\bsertitle{Society of Photo-Optical Instrumentation Engineers (SPIE) Conference
  Series},
vol. \bseriesno{4497},
pp. \bfpage{115}--\blpage{126}
(\byear{2002}).
\doiurl{10.1117/12.454217}
\end{bchapter}
\endbibitem

\bibitem[\protect\citeauthoryear{{Willingale}
  et~al.}{2016}]{2016SPIE.9905E..1YW}
\begin{bchapter}
\bauthor{\bsnm{{Willingale}}, \binits{R.}},
\bauthor{\bsnm{{Pearson}}, \binits{J.F.}},
\bauthor{\bsnm{{Martindale}}, \binits{A.}},
\bauthor{\bsnm{{Feldman}}, \binits{C.H.}},
\bauthor{\bsnm{{Fairbend}}, \binits{R.}},
\bauthor{\bsnm{{Schyns}}, \binits{E.}},
\bauthor{\bsnm{{Petit}}, \binits{S.}},
\bauthor{\bsnm{{Osborne}}, \binits{J.P.}},
\bauthor{\bsnm{{O'Brien}}, \binits{P.T.}}:
\bctitle{{Aberrations in square pore micro-channel optics used for x-ray
  lobster eye telescopes}}.
In: \beditor{\bsnm{{den Herder}}, \binits{J.-W.A.}},
\beditor{\bsnm{{Takahashi}}, \binits{T.}},
\beditor{\bsnm{{Bautz}}, \binits{M.}} (eds.)
\bbtitle{Space Telescopes and Instrumentation 2016: Ultraviolet to Gamma Ray}.
\bsertitle{Society of Photo-Optical Instrumentation Engineers (SPIE) Conference
  Series},
vol. \bseriesno{9905},
p. \bfpage{99051}
(\byear{2016}).
\doiurl{10.1117/12.2232946}
\end{bchapter}
\endbibitem

\bibitem[\protect\citeauthoryear{{Ling} et~al.}{2023}]{2023LingZXRAA}
\begin{barticle}
\bauthor{\bsnm{{Ling}}, \binits{Z.X.}},
\bauthor{\bsnm{{Sun}}, \binits{X.J.}},
\bauthor{\bsnm{{Zhang}}, \binits{C.}},
\bauthor{\bsnm{{Sun}}, \binits{S.L.}},
\bauthor{\bsnm{{Jin}}, \binits{G.}},
\bauthor{\bsnm{{Zhang}}, \binits{S.N.}},
\bauthor{\bsnm{{Zhang}}, \binits{X.F.}},
\bauthor{\bsnm{{Chang}}, \binits{J.B.}},
\bauthor{\bsnm{{Chen}}, \binits{F.S.}},
\bauthor{\bsnm{{Chen}}, \binits{Y.F.}},
\bauthor{\bsnm{{Cheng}}, \binits{Z.W.}},
\bauthor{\bsnm{{Fu}}, \binits{W.}},
\bauthor{\bsnm{{Han}}, \binits{Y.X.}},
\bauthor{\bsnm{{Li}}, \binits{H.}},
\bauthor{\bsnm{{Li}}, \binits{J.F.}},
\bauthor{\bsnm{{Li}}, \binits{Y.}},
\bauthor{\bsnm{{Li}}, \binits{Z.D.}},
\bauthor{\bsnm{{Liu}}, \binits{P.R.}},
\bauthor{\bsnm{{Lv}}, \binits{Y.H.}},
\bauthor{\bsnm{{Ma}}, \binits{X.H.}},
\bauthor{\bsnm{{Tang}}, \binits{Y.J.}},
\bauthor{\bsnm{{Wang}}, \binits{C.B.}},
\bauthor{\bsnm{{Xie}}, \binits{R.J.}},
\bauthor{\bsnm{{Xue}}, \binits{Y.L.}},
\bauthor{\bsnm{{Yan}}, \binits{A.L.}},
\bauthor{\bsnm{{Zhang}}, \binits{Q.}},
\bauthor{\bsnm{{Bao}}, \binits{C.Y.}},
\bauthor{\bsnm{{Cai}}, \binits{H.B.}},
\bauthor{\bsnm{{Cheng}}, \binits{H.Q.}},
\bauthor{\bsnm{{Cui}}, \binits{C.Z.}},
\bauthor{\bsnm{{Dai}}, \binits{Y.F.}},
\bauthor{\bsnm{{Fan}}, \binits{D.W.}},
\bauthor{\bsnm{{Hu}}, \binits{H.B.}},
\bauthor{\bsnm{{Hu}}, \binits{J.W.}},
\bauthor{\bsnm{{Huang}}, \binits{M.H.}},
\bauthor{\bsnm{{Jia}}, \binits{Z.Q.}},
\bauthor{\bsnm{{Jin}}, \binits{C.C.}},
\bauthor{\bsnm{{Li}}, \binits{D.Y.}},
\bauthor{\bsnm{{Li}}, \binits{J.Q.}},
\bauthor{\bsnm{{Liu}}, \binits{H.Y.}},
\bauthor{\bsnm{{Liu}}, \binits{M.J.}},
\bauthor{\bsnm{{Liu}}, \binits{Y.}},
\bauthor{\bsnm{{Pan}}, \binits{H.W.}},
\bauthor{\bsnm{{Qiu}}, \binits{Y.L.}},
\bauthor{\bsnm{{Sugizaki}}, \binits{M.}},
\bauthor{\bsnm{{Sun}}, \binits{H.}},
\bauthor{\bsnm{{Wang}}, \binits{W.X.}},
\bauthor{\bsnm{{Wang}}, \binits{Y.L.}},
\bauthor{\bsnm{{Wu}}, \binits{Q.Y.}},
\bauthor{\bsnm{{Xu}}, \binits{X.P.}},
\bauthor{\bsnm{{Xu}}, \binits{Y.F.}},
\bauthor{\bsnm{{Yang}}, \binits{H.N.}},
\bauthor{\bsnm{{Yang}}, \binits{X.}},
\bauthor{\bsnm{{Zhang}}, \binits{B.}},
\bauthor{\bsnm{{Zhang}}, \binits{M.}},
\bauthor{\bsnm{{Zhang}}, \binits{W.D.}},
\bauthor{\bsnm{{Zhang}}, \binits{Z.}},
\bauthor{\bsnm{{Zhao}}, \binits{D.H.}},
\bauthor{\bsnm{{Cong}}, \binits{X.Q.}},
\bauthor{\bsnm{{Jiang}}, \binits{B.W.}},
\bauthor{\bsnm{{Li}}, \binits{L.H.}},
\bauthor{\bsnm{{Qiu}}, \binits{X.B.}},
\bauthor{\bsnm{{Sun}}, \binits{J.N.}},
\bauthor{\bsnm{{Su}}, \binits{D.T.}},
\bauthor{\bsnm{{Wang}}, \binits{J.}},
\bauthor{\bsnm{{Wu}}, \binits{C.}},
\bauthor{\bsnm{{Xu}}, \binits{Z.}},
\bauthor{\bsnm{{Yang}}, \binits{X.M.}},
\bauthor{\bsnm{{Zhang}}, \binits{S.K.}},
\bauthor{\bsnm{{Zhang}}, \binits{Z.}},
\bauthor{\bsnm{{Zhang}}, \binits{N.}},
\bauthor{\bsnm{{Zhu}}, \binits{Y.F.}},
\bauthor{\bsnm{{Ban}}, \binits{H.Y.}},
\bauthor{\bsnm{{Bi}}, \binits{X.Z.}},
\bauthor{\bsnm{{Cai}}, \binits{Z.M.}},
\bauthor{\bsnm{{Chen}}, \binits{W.}},
\bauthor{\bsnm{{Chen}}, \binits{X.}},
\bauthor{\bsnm{{Chen}}, \binits{Y.H.}},
\bauthor{\bsnm{{Cui}}, \binits{Y.}},
\bauthor{\bsnm{{Duan}}, \binits{X.L.}},
\bauthor{\bsnm{{Feng}}, \binits{Z.G.}},
\bauthor{\bsnm{{Gao}}, \binits{Y.}},
\bauthor{\bsnm{{He}}, \binits{J.W.}},
\bauthor{\bsnm{{He}}, \binits{T.}},
\bauthor{\bsnm{{Huang}}, \binits{J.J.}},
\bauthor{\bsnm{{Li}}, \binits{F.}},
\bauthor{\bsnm{{Li}}, \binits{J.S.}},
\bauthor{\bsnm{{Li}}, \binits{T.J.}},
\bauthor{\bsnm{{Li}}, \binits{T.T.}},
\bauthor{\bsnm{{Liu}}, \binits{H.Q.}},
\bauthor{\bsnm{{Liu}}, \binits{L.}},
\bauthor{\bsnm{{Liu}}, \binits{R.}},
\bauthor{\bsnm{{Liu}}, \binits{S.}},
\bauthor{\bsnm{{Meng}}, \binits{N.}},
\bauthor{\bsnm{{Shi}}, \binits{Q.}},
\bauthor{\bsnm{{Sun}}, \binits{A.T.}},
\bauthor{\bsnm{{Wang}}, \binits{Y.M.}},
\bauthor{\bsnm{{Wang}}, \binits{Y.B.}},
\bauthor{\bsnm{{Wu}}, \binits{H.C.}},
\bauthor{\bsnm{{Xu}}, \binits{D.X.}},
\bauthor{\bsnm{{Yang}}, \binits{Y.Q.}},
\bauthor{\bsnm{{Yang}}, \binits{Y.}},
\bauthor{\bsnm{{Yu}}, \binits{X.S.}},
\bauthor{\bsnm{{Zhang}}, \binits{K.X.}},
\bauthor{\bsnm{{Zhang}}, \binits{Y.L.}},
\bauthor{\bsnm{{Zhang}}, \binits{Y.H.}},
\bauthor{\bsnm{{Zhang}}, \binits{Y.T.}},
\bauthor{\bsnm{{Zhou}}, \binits{H.}},
\bauthor{\bsnm{{Zhu}}, \binits{X.C.}},
\bauthor{\bsnm{{Cheng}}, \binits{J.S.}},
\bauthor{\bsnm{{Qin}}, \binits{L.}},
\bauthor{\bsnm{{Wang}}, \binits{L.}},
\bauthor{\bsnm{{Wang}}, \binits{Q.L.}},
\bauthor{\bsnm{{Bai}}, \binits{M.}},
\bauthor{\bsnm{{Gao}}, \binits{R.L.}},
\bauthor{\bsnm{{Ji}}, \binits{Z.}},
\bauthor{\bsnm{{Liu}}, \binits{Y.R.}},
\bauthor{\bsnm{{Ma}}, \binits{F.L.}},
\bauthor{\bsnm{{Shi}}, \binits{Y.J.}},
\bauthor{\bsnm{{Su}}, \binits{J.}},
\bauthor{\bsnm{{Tan}}, \binits{Y.Y.}},
\bauthor{\bsnm{{Tong}}, \binits{J.Z.}},
\bauthor{\bsnm{{Xu}}, \binits{H.T.}},
\bauthor{\bsnm{{Xue}}, \binits{C.B.}},
\bauthor{\bsnm{{Xue}}, \binits{G.F.}},
\bauthor{\bsnm{{Yuan}}, \binits{W.}}:
\batitle{{The Lobster Eye Imager for Astronomy Onboard the SATech-01
  Satellite}}.
\bjtitle{Research in Astronomy and Astrophysics}
\bvolume{23}(\bissue{9}),
\bfpage{095007}
(\byear{2023})
\doiurl{10.1088/1674-4527/acd593}
{\href{https://arxiv.org/abs/2305.14895}{{arXiv:2305.14895}}}
{[astro-ph.IM]}
\end{barticle}
\endbibitem

\bibitem[\protect\citeauthoryear{{Yuan} et~al.}{2015}]{2015arXiv150607735Y}
\begin{botherref}
\oauthor{\bsnm{{Yuan}}, \binits{W.}},
\oauthor{\bsnm{{Zhang}}, \binits{C.}},
\oauthor{\bsnm{{Feng}}, \binits{H.}},
\oauthor{\bsnm{{Zhang}}, \binits{S.N.}},
\oauthor{\bsnm{{Ling}}, \binits{Z.X.}},
\oauthor{\bsnm{{Zhao}}, \binits{D.}},
\oauthor{\bsnm{{Deng}}, \binits{J.}},
\oauthor{\bsnm{{Qiu}}, \binits{Y.}},
\oauthor{\bsnm{{Osborne}}, \binits{J.P.}},
\oauthor{\bsnm{{O'Brien}}, \binits{P.}},
\oauthor{\bsnm{{Willingale}}, \binits{R.}},
\oauthor{\bsnm{{Lapington}}, \binits{J.}},
\oauthor{\bsnm{{Fraser}}, \binits{G.W.}},
\oauthor{\bsnm{{the Einstein Probe team}}}:
{Einstein Probe - a small mission to monitor and explore the dynamic X-ray
  Universe}.
arXiv e-prints,
1506--07735
(2015)
{\href{https://arxiv.org/abs/1506.07735}{{arXiv:1506.07735}}}
{[astro-ph.HE]}
\end{botherref}
\endbibitem

\bibitem[\protect\citeauthoryear{{Yuan} et~al.}{2018}]{2018SPIE10699E..25Y}
\begin{bchapter}
\bauthor{\bsnm{{Yuan}}, \binits{W.}},
\bauthor{\bsnm{{Zhang}}, \binits{C.}},
\bauthor{\bsnm{{Ling}}, \binits{Z.}},
\bauthor{\bsnm{{Zhao}}, \binits{D.}},
\bauthor{\bsnm{{Wang}}, \binits{W.}},
\bauthor{\bsnm{{Chen}}, \binits{Y.}},
\bauthor{\bsnm{{Lu}}, \binits{F.}},
\bauthor{\bsnm{{Zhang}}, \binits{S.-N.}},
\bauthor{\bsnm{{Cui}}, \binits{W.}}:
\bctitle{{Einstein Probe: a lobster-eye telescope for monitoring the x-ray
  sky}}.
In: \beditor{\bsnm{{den Herder}}, \binits{J.-W.A.}},
\beditor{\bsnm{{Nikzad}}, \binits{S.}},
\beditor{\bsnm{{Nakazawa}}, \binits{K.}} (eds.)
\bbtitle{Space Telescopes and Instrumentation 2018: Ultraviolet to Gamma Ray}.
\bsertitle{Society of Photo-Optical Instrumentation Engineers (SPIE) Conference
  Series},
vol. \bseriesno{10699},
p. \bfpage{1069925}
(\byear{2018}).
\doiurl{10.1117/12.2313358}
\end{bchapter}
\endbibitem

\bibitem[\protect\citeauthoryear{{Yuan} et~al.}{2022}]{2022hxga.book...86Y}
\begin{bchapter}
\bauthor{\bsnm{{Yuan}}, \binits{W.}},
\bauthor{\bsnm{{Zhang}}, \binits{C.}},
\bauthor{\bsnm{{Chen}}, \binits{Y.}},
\bauthor{\bsnm{{Ling}}, \binits{Z.}}:
\bctitle{{The Einstein Probe Mission}}.
In: \bbtitle{Handbook of X-ray and Gamma-ray Astrophysics},
p. \bfpage{86}
(\byear{2022}).
\doiurl{10.1007/978-981-16-4544-0_151-1}
\end{bchapter}
\endbibitem

\bibitem[\protect\citeauthoryear{{Zhang} et~al.}{2022}]{2022ZhangChenApJL}
\begin{barticle}
\bauthor{\bsnm{{Zhang}}, \binits{C.}},
\bauthor{\bsnm{{Ling}}, \binits{Z.X.}},
\bauthor{\bsnm{{Sun}}, \binits{X.J.}},
\bauthor{\bsnm{{Sun}}, \binits{S.L.}},
\bauthor{\bsnm{{Liu}}, \binits{Y.}},
\bauthor{\bsnm{{Li}}, \binits{Z.D.}},
\bauthor{\bsnm{{Xue}}, \binits{Y.L.}},
\bauthor{\bsnm{{Chen}}, \binits{Y.F.}},
\bauthor{\bsnm{{Dai}}, \binits{Y.F.}},
\bauthor{\bsnm{{Jia}}, \binits{Z.Q.}},
\bauthor{\bsnm{{Liu}}, \binits{H.Y.}},
\bauthor{\bsnm{{Zhang}}, \binits{X.F.}},
\bauthor{\bsnm{{Zhang}}, \binits{Y.H.}},
\bauthor{\bsnm{{Zhang}}, \binits{S.N.}},
\bauthor{\bsnm{{Chen}}, \binits{F.S.}},
\bauthor{\bsnm{{Cheng}}, \binits{Z.W.}},
\bauthor{\bsnm{{Fu}}, \binits{W.}},
\bauthor{\bsnm{{Han}}, \binits{Y.X.}},
\bauthor{\bsnm{{Li}}, \binits{H.}},
\bauthor{\bsnm{{Li}}, \binits{J.F.}},
\bauthor{\bsnm{{Li}}, \binits{Y.}},
\bauthor{\bsnm{{Liu}}, \binits{P.R.}},
\bauthor{\bsnm{{Ma}}, \binits{X.H.}},
\bauthor{\bsnm{{Tang}}, \binits{Y.J.}},
\bauthor{\bsnm{{Wang}}, \binits{C.B.}},
\bauthor{\bsnm{{Xie}}, \binits{R.J.}},
\bauthor{\bsnm{{Yan}}, \binits{A.L.}},
\bauthor{\bsnm{{Zhang}}, \binits{Q.}},
\bauthor{\bsnm{{Jiang}}, \binits{B.W.}},
\bauthor{\bsnm{{Jin}}, \binits{G.}},
\bauthor{\bsnm{{Li}}, \binits{L.H.}},
\bauthor{\bsnm{{Qiu}}, \binits{X.B.}},
\bauthor{\bsnm{{Su}}, \binits{D.T.}},
\bauthor{\bsnm{{Sun}}, \binits{J.N.}},
\bauthor{\bsnm{{Xu}}, \binits{Z.}},
\bauthor{\bsnm{{Zhang}}, \binits{S.K.}},
\bauthor{\bsnm{{Zhang}}, \binits{Z.}},
\bauthor{\bsnm{{Zhang}}, \binits{N.}},
\bauthor{\bsnm{{Bi}}, \binits{X.Z.}},
\bauthor{\bsnm{{Cai}}, \binits{Z.M.}},
\bauthor{\bsnm{{He}}, \binits{J.W.}},
\bauthor{\bsnm{{Liu}}, \binits{H.Q.}},
\bauthor{\bsnm{{Zhu}}, \binits{X.C.}},
\bauthor{\bsnm{{Cheng}}, \binits{H.Q.}},
\bauthor{\bsnm{{Cui}}, \binits{C.Z.}},
\bauthor{\bsnm{{Fan}}, \binits{D.W.}},
\bauthor{\bsnm{{Hu}}, \binits{H.B.}},
\bauthor{\bsnm{{Huang}}, \binits{M.H.}},
\bauthor{\bsnm{{Jin}}, \binits{C.C.}},
\bauthor{\bsnm{{Li}}, \binits{D.Y.}},
\bauthor{\bsnm{{Pan}}, \binits{H.W.}},
\bauthor{\bsnm{{Wang}}, \binits{W.X.}},
\bauthor{\bsnm{{Xu}}, \binits{Y.F.}},
\bauthor{\bsnm{{Yang}}, \binits{X.}},
\bauthor{\bsnm{{Zhang}}, \binits{B.}},
\bauthor{\bsnm{{Zhang}}, \binits{M.}},
\bauthor{\bsnm{{Zhang}}, \binits{W.D.}},
\bauthor{\bsnm{{Zhao}}, \binits{D.H.}},
\bauthor{\bsnm{{Bai}}, \binits{M.}},
\bauthor{\bsnm{{Ji}}, \binits{Z.}},
\bauthor{\bsnm{{Liu}}, \binits{Y.R.}},
\bauthor{\bsnm{{Ma}}, \binits{F.L.}},
\bauthor{\bsnm{{Su}}, \binits{J.}},
\bauthor{\bsnm{{Tong}}, \binits{J.Z.}},
\bauthor{\bsnm{{Wang}}, \binits{Y.S.}},
\bauthor{\bsnm{{Zhao}}, \binits{Z.J.}},
\bauthor{\bsnm{{Feldman}}, \binits{C.}},
\bauthor{\bsnm{{O'Brien}}, \binits{P.}},
\bauthor{\bsnm{{Osborne}}, \binits{J.P.}},
\bauthor{\bsnm{{Willingale}}, \binits{R.}},
\bauthor{\bsnm{{Burwitz}}, \binits{V.}},
\bauthor{\bsnm{{Hartner}}, \binits{G.}},
\bauthor{\bsnm{{Langmeier}}, \binits{A.}},
\bauthor{\bsnm{{M{\"u}ller}}, \binits{T.}},
\bauthor{\bsnm{{Rukdee}}, \binits{S.}},
\bauthor{\bsnm{{Schmidt}}, \binits{T.}},
\bauthor{\bsnm{{Kuulkers}}, \binits{E.}},
\bauthor{\bsnm{{Yuan}}, \binits{W.}}:
\batitle{{First Wide Field-of-view X-Ray Observations by a Lobster-eye Focusing
  Telescope in Orbit}}.
\bjtitle{\apjl}
\bvolume{941}(\bissue{1}),
\bfpage{2}
(\byear{2022})
\doiurl{10.3847/2041-8213/aca32f}
{\href{https://arxiv.org/abs/2211.10007}{{arXiv:2211.10007}}}
{[astro-ph.HE]}
\end{barticle}
\endbibitem

\bibitem[\protect\citeauthoryear{{Wu} et~al.}{2022}]{2022PASP..134c5006W}
\begin{barticle}
\bauthor{\bsnm{{Wu}}, \binits{Q.}},
\bauthor{\bsnm{{Jia}}, \binits{Z.}},
\bauthor{\bsnm{{Wang}}, \binits{W.}},
\bauthor{\bsnm{{Ling}}, \binits{Z.}},
\bauthor{\bsnm{{Zhang}}, \binits{C.}},
\bauthor{\bsnm{{Zhang}}, \binits{S.}},
\bauthor{\bsnm{{Yuan}}, \binits{W.}}:
\batitle{{X-Ray Performance of a Customized Large-format Scientific CMOS
  Detector}}.
\bjtitle{\pasp}
\bvolume{134}(\bissue{1033}),
\bfpage{035006}
(\byear{2022})
\doiurl{10.1088/1538-3873/ac5ac9}
\end{barticle}
\endbibitem

\bibitem[\protect\citeauthoryear{Wu et~al.}{2023}]{wu2023}
\begin{barticle}
\bauthor{\bsnm{Wu}, \binits{Q.}},
\bauthor{\bsnm{Ling}, \binits{Z.}},
\bauthor{\bsnm{Zhang}, \binits{C.}},
\bauthor{\bsnm{Zhang}, \binits{S.-N.}},
\bauthor{\bsnm{Yuan}, \binits{W.}}:
\batitle{An aluminum-coated scmos sensor for x-ray astronomy}.
\bjtitle{Publications of the Astronomical Society of the Pacific}
\bvolume{135}(\bissue{1053}),
\bfpage{115002}
(\byear{2023})
\doiurl{10.1088/1538-3873/ad03d7}
\end{barticle}
\endbibitem

\bibitem[\protect\citeauthoryear{{Zhang} et~al.}{2012}]{2012SPIE.8443E..3XZ}
\begin{bchapter}
\bauthor{\bsnm{{Zhang}}, \binits{C.}},
\bauthor{\bsnm{{Ling}}, \binits{Z.}},
\bauthor{\bsnm{{Zhang}}, \binits{S.-N.}}:
\bctitle{{Development of the super-high angular resolution principle for x-ray
  imaging: experimental demonstrations}}.
In: \beditor{\bsnm{{Takahashi}}, \binits{T.}},
\beditor{\bsnm{{Murray}}, \binits{S.S.}},
\beditor{\bsnm{{den Herder}}, \binits{J.-W.A.}} (eds.)
\bbtitle{Space Telescopes and Instrumentation 2012: Ultraviolet to Gamma Ray}.
\bsertitle{Society of Photo-Optical Instrumentation Engineers (SPIE) Conference
  Series},
vol. \bseriesno{8443},
p. \bfpage{84433}
(\byear{2012}).
\doiurl{10.1117/12.925780}
\end{bchapter}
\endbibitem

\bibitem[\protect\citeauthoryear{{Freyberg} et~al.}{2005}]{2005ExA....20..405F}
\begin{barticle}
\bauthor{\bsnm{{Freyberg}}, \binits{M.J.}},
\bauthor{\bsnm{{Br{\"a}uninger}}, \binits{H.}},
\bauthor{\bsnm{{Burkert}}, \binits{W.}},
\bauthor{\bsnm{{Hartner}}, \binits{G.D.}},
\bauthor{\bsnm{{Citterio}}, \binits{O.}},
\bauthor{\bsnm{{Mazzoleni}}, \binits{F.}},
\bauthor{\bsnm{{Pareschi}}, \binits{G.}},
\bauthor{\bsnm{{Spiga}}, \binits{D.}},
\bauthor{\bsnm{{Romaine}}, \binits{S.}},
\bauthor{\bsnm{{Gorenstein}}, \binits{P.}},
\bauthor{\bsnm{{Ramsey}}, \binits{B.D.}}:
\batitle{{The MPE X-ray test facility PANTER: Calibration of hard X-ray (15-50
  kev) optics}}.
\bjtitle{Experimental Astronomy}
\bvolume{20}(\bissue{1-3}),
\bfpage{405}--\blpage{412}
(\byear{2005})
\doiurl{10.1007/s10686-006-9068-8}
\end{barticle}
\endbibitem

\bibitem[\protect\citeauthoryear{{Bradshaw} et~al.}{2019}]{2019SPIE11119E..16B}
\begin{bchapter}
\bauthor{\bsnm{{Bradshaw}}, \binits{M.}},
\bauthor{\bsnm{{Burwitz}}, \binits{V.}},
\bauthor{\bsnm{{Hartner}}, \binits{G.}},
\bauthor{\bsnm{{Pelliciari}}, \binits{C.}},
\bauthor{\bsnm{{Langmeier}}, \binits{A.}},
\bauthor{\bsnm{{Liao}}, \binits{Y.}},
\bauthor{\bsnm{{Friedrich}}, \binits{P.}},
\bauthor{\bsnm{{Valsecchi}}, \binits{G.}},
\bauthor{\bsnm{{Barri{\`e}re}}, \binits{N.}},
\bauthor{\bsnm{{Collon}}, \binits{M.J.}},
\bauthor{\bsnm{{Vacanti}}, \binits{G.}}:
\bctitle{{Developments in testing x-ray optics at MPE's PANTER facility}}.
In: \beditor{\bsnm{{O'Dell}}, \binits{S.L.}},
\beditor{\bsnm{{Pareschi}}, \binits{G.}} (eds.)
\bbtitle{Optics for EUV, X-Ray, and Gamma-Ray Astronomy IX}.
\bsertitle{Society of Photo-Optical Instrumentation Engineers (SPIE) Conference
  Series},
vol. \bseriesno{11119},
p. \bfpage{1111916}
(\byear{2019}).
\doiurl{10.1117/12.2531709}
\end{bchapter}
\endbibitem

\bibitem[\protect\citeauthoryear{{Rukdee} et~al.}{2023}]{2023SPIE12777E..3FR}
\begin{bchapter}
\bauthor{\bsnm{{Rukdee}}, \binits{S.}},
\bauthor{\bsnm{{Burwitz}}, \binits{V.}},
\bauthor{\bsnm{{Hartner}}, \binits{G.}},
\bauthor{\bsnm{{Mu{\`I}ller}}, \binits{T.}},
\bauthor{\bsnm{{Schmidt}}, \binits{T.}},
\bauthor{\bsnm{{Langmeier}}, \binits{A.}},
\bauthor{\bsnm{{Friedrich}}, \binits{P.}},
\bauthor{\bsnm{{Feldman}}, \binits{C.}},
\bauthor{\bsnm{{O'Brien}}, \binits{P.}},
\bauthor{\bsnm{{Willingale}}, \binits{R.}},
\bauthor{\bsnm{{Zhang}}, \binits{C.}},
\bauthor{\bsnm{{Ling}}, \binits{Z.}},
\bauthor{\bsnm{{Yuan}}, \binits{W.}}:
\bctitle{{The X-ray testing of Einstein Probe Wide-field X-ray Telescope
  Qualification Model at PANTER}}.
In: \beditor{\bsnm{{Minoglou}}, \binits{K.}},
\beditor{\bsnm{{Karafolas}}, \binits{N.}},
\beditor{\bsnm{{Cugny}}, \binits{B.}} (eds.)
\bbtitle{Society of Photo-Optical Instrumentation Engineers (SPIE) Conference
  Series}.
\bsertitle{Society of Photo-Optical Instrumentation Engineers (SPIE) Conference
  Series},
vol. \bseriesno{12777},
p. \bfpage{127773}
(\byear{2023}).
\doiurl{10.1117/12.2690354}
\end{bchapter}
\endbibitem

\bibitem[\protect\citeauthoryear{{Wang} et~al.}{2023}]{2023ExA....55..427W}
\begin{barticle}
\bauthor{\bsnm{{Wang}}, \binits{Y.}},
\bauthor{\bsnm{{Zhao}}, \binits{Z.}},
\bauthor{\bsnm{{Hou}}, \binits{D.}},
\bauthor{\bsnm{{Yang}}, \binits{X.}},
\bauthor{\bsnm{{Chen}}, \binits{C.}},
\bauthor{\bsnm{{Li}}, \binits{X.}},
\bauthor{\bsnm{{Zhu}}, \binits{Y.}},
\bauthor{\bsnm{{Zhao}}, \binits{X.}},
\bauthor{\bsnm{{Ma}}, \binits{J.}},
\bauthor{\bsnm{{Xu}}, \binits{H.}},
\bauthor{\bsnm{{Chen}}, \binits{Y.}},
\bauthor{\bsnm{{Wang}}, \binits{G.}},
\bauthor{\bsnm{{Lu}}, \binits{F.}},
\bauthor{\bsnm{{Zhang}}, \binits{S.}},
\bauthor{\bsnm{{Zhang}}, \binits{S.}},
\bauthor{\bsnm{{Chen}}, \binits{Y.}},
\bauthor{\bsnm{{Xu}}, \binits{Y.}}:
\batitle{{The 100-m X-ray test facility at IHEP}}.
\bjtitle{Experimental Astronomy}
\bvolume{55}(\bissue{2}),
\bfpage{427}--\blpage{445}
(\byear{2023})
\doiurl{10.1007/s10686-022-09872-7}
{\href{https://arxiv.org/abs/2210.15840}{{arXiv:2210.15840}}}
{[astro-ph.IM]}
\end{barticle}
\endbibitem

\bibitem[\protect\citeauthoryear{{Zhao} et~al.}{2014}]{2014SPIE.9144E..4EZ}
\begin{bchapter}
\bauthor{\bsnm{{Zhao}}, \binits{D.}},
\bauthor{\bsnm{{Zhang}}, \binits{C.}},
\bauthor{\bsnm{{Yuan}}, \binits{W.}},
\bauthor{\bsnm{{Willingale}}, \binits{R.}},
\bauthor{\bsnm{{Ling}}, \binits{Z.}},
\bauthor{\bsnm{{Feng}}, \binits{H.}},
\bauthor{\bsnm{{Li}}, \binits{H.}},
\bauthor{\bsnm{{Ji}}, \binits{J.}},
\bauthor{\bsnm{{Wang}}, \binits{W.}},
\bauthor{\bsnm{{Zhang}}, \binits{S.}}:
\bctitle{{Ray tracing simulations for the wide-field x-ray telescope of the
  Einstein Probe mission based on Geant4 and XRTG4}}.
In: \beditor{\bsnm{{Takahashi}}, \binits{T.}},
\beditor{\bsnm{{den Herder}}, \binits{J.-W.A.}},
\beditor{\bsnm{{Bautz}}, \binits{M.}} (eds.)
\bbtitle{Space Telescopes and Instrumentation 2014: Ultraviolet to Gamma Ray}.
\bsertitle{Society of Photo-Optical Instrumentation Engineers (SPIE) Conference
  Series},
vol. \bseriesno{9144},
p. \bfpage{91444}
(\byear{2014}).
\doiurl{10.1117/12.2055434}
\end{bchapter}
\endbibitem

\bibitem[\protect\citeauthoryear{{Li} et~al.}{2021}]{2021OptCo.48326656L}
\begin{barticle}
\bauthor{\bsnm{{Li}}, \binits{L.}},
\bauthor{\bsnm{{Zhang}}, \binits{C.}},
\bauthor{\bsnm{{Jin}}, \binits{G.}},
\bauthor{\bsnm{{Yuan}}, \binits{W.}},
\bauthor{\bsnm{{Zhang}}, \binits{S.}},
\bauthor{\bsnm{{Li}}, \binits{Z.}},
\bauthor{\bsnm{{Gu}}, \binits{Y.}},
\bauthor{\bsnm{{Wang}}, \binits{J.}},
\bauthor{\bsnm{{Zhang}}, \binits{Z.}},
\bauthor{\bsnm{{Zhang}}, \binits{Z.}},
\bauthor{\bsnm{{Xu}}, \binits{Z.}},
\bauthor{\bsnm{{Jiang}}, \binits{B.}},
\bauthor{\bsnm{{Wu}}, \binits{C.}},
\bauthor{\bsnm{{Liao}}, \binits{D.}},
\bauthor{\bsnm{{Ling}}, \binits{Z.}},
\bauthor{\bsnm{{Zhao}}, \binits{D.}}:
\batitle{{Study on the optical properties of Angel Lobster eye X-ray flat micro
  pore optical device}}.
\bjtitle{Optics Communications}
\bvolume{483},
\bfpage{126656}
(\byear{2021})
\doiurl{10.1016/j.optcom.2020.126656}
\end{barticle}
\endbibitem

\bibitem[\protect\citeauthoryear{{Li} et~al.}{2022}]{2022PASP..134k5002L}
\begin{barticle}
\bauthor{\bsnm{{Li}}, \binits{L.}},
\bauthor{\bsnm{{Zhang}}, \binits{Y.}},
\bauthor{\bsnm{{Ouyang}}, \binits{M.}},
\bauthor{\bsnm{{Wang}}, \binits{J.}},
\bauthor{\bsnm{{Zhang}}, \binits{Z.}},
\bauthor{\bsnm{{Zhang}}, \binits{C.}},
\bauthor{\bsnm{{Ling}}, \binits{Z.}},
\bauthor{\bsnm{{Fang}}, \binits{J.}},
\bauthor{\bsnm{{Fu}}, \binits{Y.}}:
\batitle{{Fabrication and Performance of Lobster Eye X-Ray Micro Pore Optics
  with the Ultra-high Aspect Ratio}}.
\bjtitle{\pasp}
\bvolume{134}(\bissue{1041}),
\bfpage{115002}
(\byear{2022})
\doiurl{10.1088/1538-3873/ac9f6d}
\end{barticle}
\endbibitem

\bibitem[\protect\citeauthoryear{{Zhao} et~al.}{2017}]{2017ExA....43..267Z}
\begin{barticle}
\bauthor{\bsnm{{Zhao}}, \binits{D.}},
\bauthor{\bsnm{{Zhang}}, \binits{C.}},
\bauthor{\bsnm{{Yuan}}, \binits{W.}},
\bauthor{\bsnm{{Zhang}}, \binits{S.}},
\bauthor{\bsnm{{Willingale}}, \binits{R.}},
\bauthor{\bsnm{{Ling}}, \binits{Z.}}:
\batitle{{Geant4 simulations of a wide-angle x-ray focusing telescope}}.
\bjtitle{Experimental Astronomy}
\bvolume{43}(\bissue{3}),
\bfpage{267}--\blpage{283}
(\byear{2017})
\doiurl{10.1007/s10686-017-9534-5}
{\href{https://arxiv.org/abs/1703.09380}{{arXiv:1703.09380}}}
{[astro-ph.IM]}
\end{barticle}
\endbibitem

\bibitem[\protect\citeauthoryear{{Feldman} et~al.}{2020}]{2020charly_wxt_LU}
\begin{bchapter}
\bauthor{\bsnm{{Feldman}}, \binits{C.}},
\bauthor{\bsnm{{O'Brien}}, \binits{P.}},
\bauthor{\bsnm{{Willingale}}, \binits{R.}},
\bauthor{\bsnm{{Zhang}}, \binits{C.}},
\bauthor{\bsnm{{Ling}}, \binits{Z.}},
\bauthor{\bsnm{{Yuan}}, \binits{W.}},
\bauthor{\bsnm{{Jia}}, \binits{Z.}},
\bauthor{\bsnm{{Jin}}, \binits{G.}},
\bauthor{\bsnm{{Li}}, \binits{L.}},
\bauthor{\bsnm{{Xu}}, \binits{Z.}},
\bauthor{\bsnm{{Zhang}}, \binits{Z.}},
\bauthor{\bsnm{{Lerman}}, \binits{H.}},
\bauthor{\bsnm{{Hutchinson}}, \binits{I.}},
\bauthor{\bsnm{{McHugh}}, \binits{M.}},
\bauthor{\bsnm{{Lodge}}, \binits{A.}}:
\bctitle{{Testing of the WXT optics at the University of Leicester}}.
In: \bbtitle{Society of Photo-Optical Instrumentation Engineers (SPIE)
  Conference Series}.
\bsertitle{Society of Photo-Optical Instrumentation Engineers (SPIE) Conference
  Series},
vol. \bseriesno{11444},
p. \bfpage{114447}
(\byear{2020}).
\doiurl{10.1117/12.2562194}
\end{bchapter}
\endbibitem

\end{thebibliography}

\end{document}